\documentclass[%
 reprint,
 amsmath,amssymb,
 aps,
]{revtex4-1}

\usepackage{graphicx}
\usepackage{float}
\usepackage{braket}
\usepackage{subcaption}
\usepackage{color, ulem}
\usepackage{gensymb}

\begin{document}

\title{Attractive electron-electron interactions from internal screening in magic angle twisted bilayer graphene}

\author{Zachary A. H. Goodwin}
\author{Fabiano Corsetti}
\author{Arash A. Mostofi}
\author{Johannes Lischner}
\affiliation{Departments of Materials and Physics and the Thomas Young Centre for Theory and Simulation of Materials, Imperial College London, South Kensington Campus, London SW7 2AZ, UK\\}

\date{\today}

\begin{abstract}
Twisted bilayer graphene (tBLG) has recently emerged as a new platform for studying electron correlations, the strength of which can be controlled via the twist angle. Here, we study the effect of internal screening on electron-electron interactions in undoped tBLG. Using the random phase approximation, we find that the dielectric response of tBLG drastically increases near the magic angle and is highly twist-angle dependent. As a consequence of the abrupt change of the Fermi velocity as a function of wave vector, the screened interaction in real space exhibits attractive regions for certain twist angles near the magic angle. Attractive interactions can induce charge density waves and superconductivity and therefore our findings could be relevant to understand the microscopic origins of the recently observed strong correlation phenomena in undoped tBLG. The resulting screened Hubbard parameters are strongly reduced and exhibit a non-linear dependence on the twist angle. We also carry out calculations with the constrained random phase approximation and parametrize a twist-angle dependent Keldysh model for the resulting effective interaction.
\end{abstract}

\maketitle

\section{Introduction}

Twistronics~\cite{TT} is the burgeoning field of controlling the electronic properties of van der Waals structures through the relative twist angle~\cite{HARA} between the component 2D materials~\cite{GBWT,MBTBLG,PTBLG,ECM,PDTBLG,ECO,EFM,NSCS,SLG,TMB,NAT_I,NAT_S,TSTBLG,SOM,SMTBLG}, with twisted bilayer graphene (tBLG) serving as the paradigmatic example~\cite{GBWT,MBTBLG,SLG,TMB,NAT_I,NAT_S,TSTBLG,SOM,SMTBLG,NAT_MEI,NAT_SS}. When the twist angle of tBLG is tuned to the magic angle (approximately 1.1\degree), the electronic bands near the Fermi level become extremely flat~\cite{LDE,GBWT,MBTBLG}, which gives rise to correlated insulating and superconducting states~\cite{NAT_S,NAT_I,TSTBLG,SOM,SMTBLG,NAT_MEI,NAT_SS}. The emergence of flat bands has recently also been found in other systems, including twisted double bilayer graphene~\cite{BIBI,FBTBB,TCT,STSCTDB} and twisted transition metal dichalcogenides~\cite{UFSS,TTMD,TITMD}. 

Many theoretical proposals have been put forward to explain the microscopic origin of the observed strong correlation phenomena in tBLG~\cite{OMACM,FTBM,ESTBLG,EDS,CMLD,LDLE,PHD_1,EPC,KDP} but, at the present time, no consensus has been achieved~\cite{EE,KVB,ECM,TJMODEL,US,CSD,KL,PS,SCHFC,IMACP,OMIB,PCIS,CTBS,SCDID,TCSSV,PMS,EPC,EPRG,MMIT,WC,NCIS,SCPKV}. Several works~\cite{SCHFC,SOM} have used Hartree-Fock theory based on a continuum model to analyze the phase diagram of tBLG. It is well known, however, that Hartree-Fock can lead to unphysical results, such as a diverging Fermi velocity in metals, because the Coulomb interaction in the exchange term is not screened. Electronic screening is also important for the construction of effective low-energy Hamiltonians as transitions between high-energy bands renormalize the interaction between flat band electrons.

The random phase approximation (RPA) is often used to describe the screened interaction between electrons. In this approach, one first determines the polarizability of non-interacting (or independent) electrons and then self-consistently computes their response to the total field consisting of the external perturbation and the induced Hartree potential of the electrons. For the construction of low-energy Hamiltonians, the constrained random phase approximation (cRPA) is used, in which transitions among low-energy bands are excluded in the polarizability. In the context of tBLG, Stauber and Kohler~\cite{PTBLG} have calculated its RPA dielectric function from a continuum model, and used it to study plasmons and collective exciton modes. More recently, Pizarro \textit{et al.}~\cite{CCRPA}, also using a continuum model, calculated the static RPA and cRPA polarizability of undoped tBLG, but only at a single twist angle of 1.05\degree. 

Here we study the cRPA and RPA screened interaction in tBLG as function of twist angle using an atomistic tight-binding model. Excluding transitions between the flat bands, we find that the cRPA screened interaction depends strongly on the twist angle and is accurately described by a Keldysh model with a twist-angle-dependent screening parameter. Inclusion of transitions between the flat bands drastically increases internal screening near the magic angle as a consequence of the emergence of flat bands. For certain twist angles near the magic angle, we find that the RPA screened interaction has \textit{attractive} regions in real space. The combination of enhanced screening and attractive regions leads to a significant reduction of the on-site Hubbard parameter, which exhibits a non-linear behaviour as function of twist angle. Finally, we discuss the implications of our findings for the phase diagram of tBLG. In particular, previous theoretical work has established that real space attractive electron-electron interactions can give rise to charge density waves and superconductivity. In this context, our discovery of attractive effective interactions in tBLG is an interesting finding that may have relevance in explaining the microscopic origin of the experimentally observed correlated insulator states and superconducting phases in tBLG.

\begin{figure*}[t!]
\begin{subfigure}{0.49\textwidth}
  \centering
  \includegraphics[width=1\linewidth]{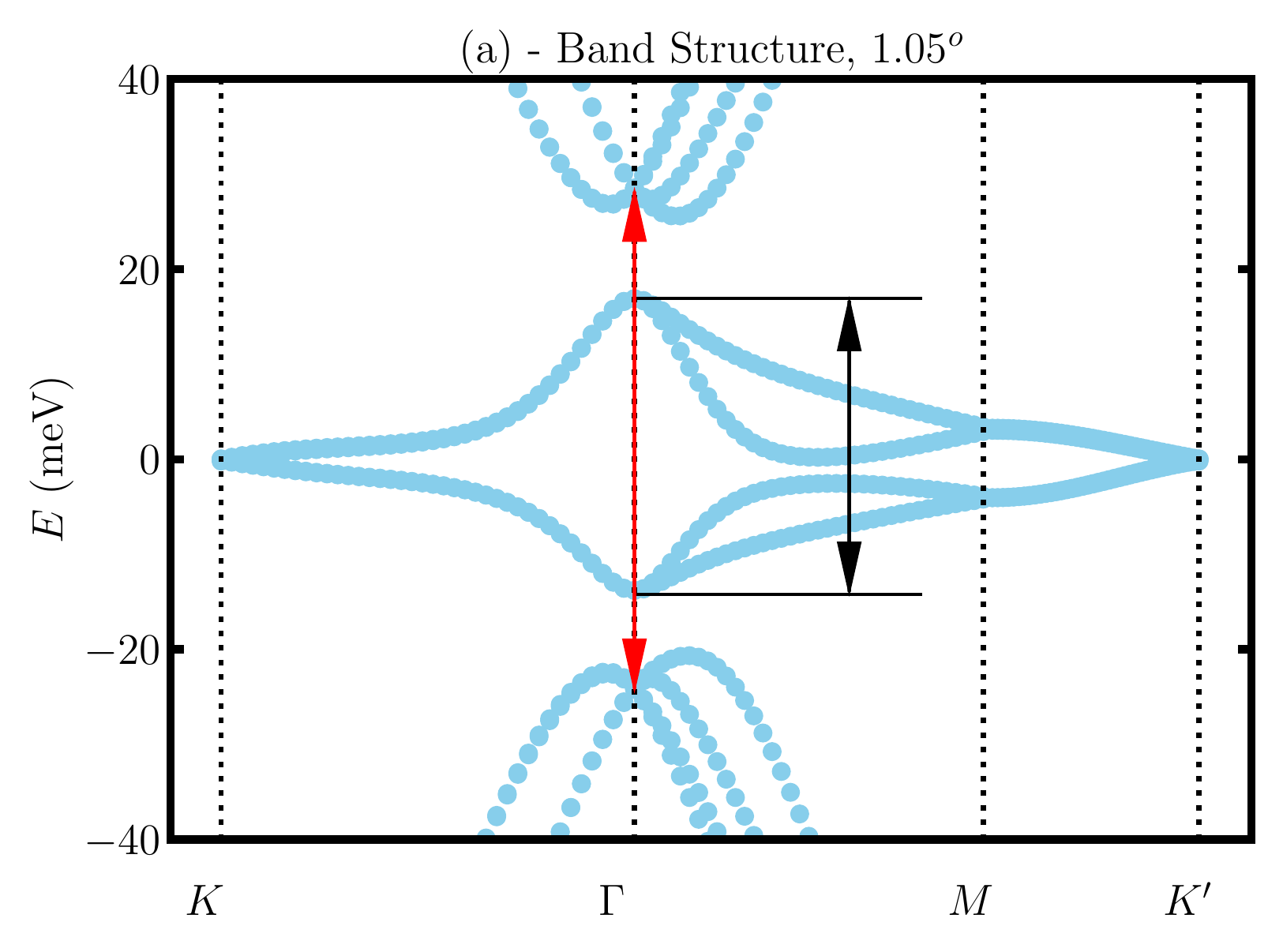}
\end{subfigure}
\begin{subfigure}{0.49\textwidth}
  \centering
  \includegraphics[width=1\linewidth]{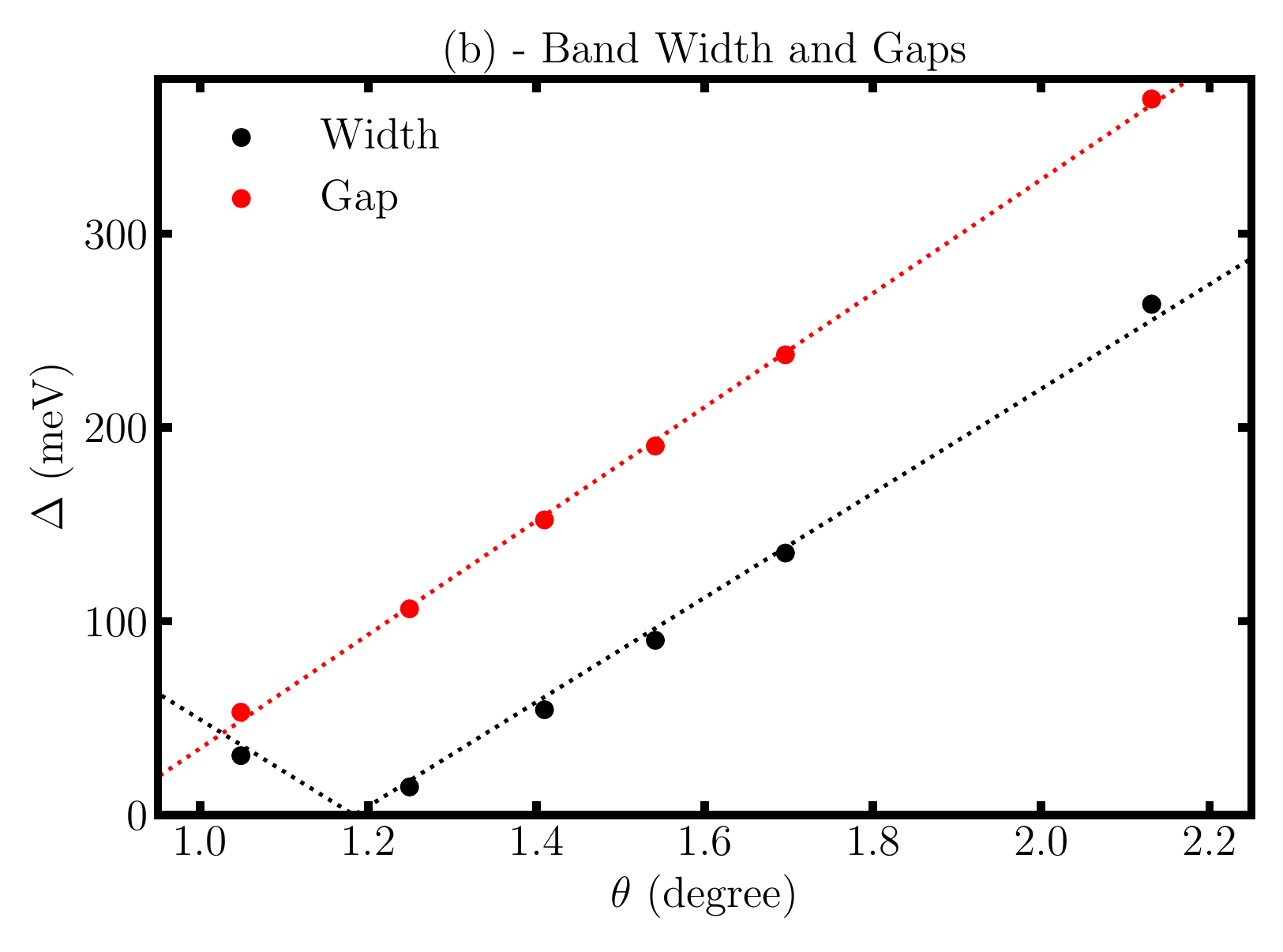}
\end{subfigure}
\caption{(a) Band structure of tBLG for a twist angle of 1.05\degree. The black arrow indicates the width of the flat bands, and the red arrow denotes the energy gap between non-flat bands at $\Gamma$. (b) Band widths of flat bands [black arrow in (a)] and band gaps at $\Gamma$ between non-flat bands [red arrow in (a)] as function of twist angle. Dotted lines are linear fits (see Appendix~\ref{ATB} for details). Figure adapted with permission from Ref.~\citenum{PHD_1}. Copyrighted by the American Physical Society.}
\label{BSBW}
\end{figure*}

\section{Methods}

\subsection{Atomistic Tight-Binding}

We use the atomistic tight-binding model of Ref.~\citenum{PHD_1} to calculate the band structure of tBLG near the  magic angle taking into account atomic corrugation (see Appendix~\ref{ATB} for details~\cite{FC}). Fig.~\ref{BSBW} shows the band structure at a twist angle of $1.05\degree$, which exhibits four flat bands near the Fermi level that are separated from all other bands by energy gaps. The undoped system is a semimetal with the flat valence and conduction bands touching at the K and K$^\prime$ points of the first Brillouin zone. As the twist angle approaches the magic angle ($\theta^*=1.18\degree$ in our calculations), the width of the flat bands decreases, see black circles in Fig.~\ref{BSBW}(b). Also, the energy gaps that separate the non-flat bands decrease as the twist angle is reduced, see red circles in Fig.~\ref{BSBW}(b). Note that in a narrow twist-angle window (1.12-1.20\degree), we find qualitatively different band structures with a metallic character for undoped tBLG (similar band structures are shown in Refs. \citenum{KDP} and \citenum{EPC}). In the rest of the paper, we only study twist angles with a semimetallic band structure (see Fig.~\ref{BS_FIG} of Appendix~\ref{ATB} for band structures). 

\subsection{Dielectric Response}

To calculate the static dielectric function of undoped tBLG, we employ the RPA. In agreement with previous work~\cite{CCRPA,PTBLG}, we find that off-diagonal elements of the dielectric matrix are small and, therefore, we focus our attention on the diagonal elements. Within these approximations~\cite{Shung,DFTRPA,SEISMG}, the dielectric function is given by 
\begin{equation}
\epsilon(\textbf{q}) = \epsilon_{\textrm{env}} + v(\textbf{q})\Pi_{0}(\textbf{q}),
\label{EPS}
\end{equation}

\noindent where $\epsilon_{\textrm{env}}$ is the environmental dielectric constant, $\textbf{q}$ is a (two-dimensional) in-plane crystal momentum, $v(\textbf{q}) = e^{2}/2\epsilon_{0}|\textbf{q}|$ is the bare Coulomb interaction in 2D and $\Pi_{0}(\textbf{q})$ is the independent-particle polarizability~\cite{Shung,DFTRPA,SEISMG}. The polarizability is obtained by evaluating the Adler-Wiser formula~\cite{Shung,DFTRPA,SEISMG} in the limit of zero temperature
\begin{equation}
\Pi_{0}(\textbf{q}) = \dfrac{4}{\Omega}\sum_{\textbf{k}}\sum_{cv}\dfrac{|\braket{\psi_{v\textbf{k}}|e^{-i\textbf{q}\cdot\textbf{r}}|\psi_{c\textbf{k}+\textbf{q}}}|^{2}}{\varepsilon_{c\textbf{k}+\textbf{q}} - \varepsilon_{v\textbf{k}}},
\label{POL}
\end{equation}

\noindent where $\Omega$ is the area of the tBLG crystal, which is proportional to the number of $k$-points in the first summation. The second summation in Eq.~\eqref{POL} is over transitions from occupied valence bands ($v$) to unoccupied conduction bands ($c$), and $\varepsilon_{c/v\textbf{k}}$ and $\psi_{c/v\textbf{k}}$ denote, respectively, the eigenvalues and Bloch states obtained from the tight-binding calculation~\cite{Shung,DFTRPA,SEISMG}. Details of the evaluation of Eq.~\eqref{POL} can be found in Appendix~\ref{AIS}. In the RPA, all transitions contribute to the summation, while in the cRPA~\cite{LEFM,CHU}, transitions between flat bands are excluded. The accuracy of the cRPA has recently been studied in Hubbard models~\cite{LcRPA} and it was found that the screening is overestimated compared to more accurate approaches. Therefore, the cRPA Hubbard parameters should be considered as lower bounds.

To calculate the cRPA polarizability, we employ a $7\times7$ regular Monkhorst-Pack  $k$-point grid to sample the Brillouin zone and sum over states that lie in an energy window $\pm 4$~eV around the Fermi energy. For the RPA, transitions between flat bands were calculated on a $35\times35$ regular Monkhorst-Pack $k$-point grid and added to the cRPA polarizability. We have found that these convergence parameters yield accurate values for the polarizability at wavevectors that do not exceed several multiples of the moir\'e reciprocal lattice vector.

\subsection{Screened Interaction and Hubbard Parameters}

The screened interaction in real space is calculated via a two-dimensional Fourier transform according to
\begin{equation}
W(\textbf{r}) = \int \dfrac{d\textbf{q}}{(2\pi)^{2}}\dfrac{v(\textbf{q})}{\epsilon(\textbf{q})}e^{-i\textbf{q}\cdot\textbf{r}}.
\label{WIR}
\end{equation}

\noindent As the polarizability is found to be approximately isotropic, the angular part of the Fourier transform can be carried out analytically and the remaining one-dimensional radial integral is done numerically (see Appendix~\ref{AIS} for details).

To determine the effective Hamiltonian of the flat band electrons, we calculate the interaction parameters of an extended Hubbard Hamiltonian via
\begin{equation}
V_{ij} = \iint d\textbf{r}d\textbf{r}^{\prime}|w_{i}(\textbf{r}^{\prime})|^{2}W(\textbf{r} - \textbf{r}^{\prime})|w_{j}(\textbf{r})|^{2}.
\label{GHI}
\end{equation}

\noindent The atomistic Wannier functions, $w_{i}(\mathbf{r})$, of the flat bands were previously constructed~\cite{MLWO,SMLWF,PHD_1} (details of which are given in Appendix~\ref{AWF}). They are centered on the AB and BA regions of the moir\'e unit cell, forming an emergent hexagonal lattice~\cite{MLWO,SMLWF,PHD_1}. They have three lobes, each of which is centered on an AA region, where the charge density of the flat bands are localized (see Fig.~\ref{VARF} of Appendix~\ref{AWF}).

\begin{figure*}[t!]
\begin{subfigure}{0.49\textwidth}
  \centering
  \includegraphics[width=1\linewidth]{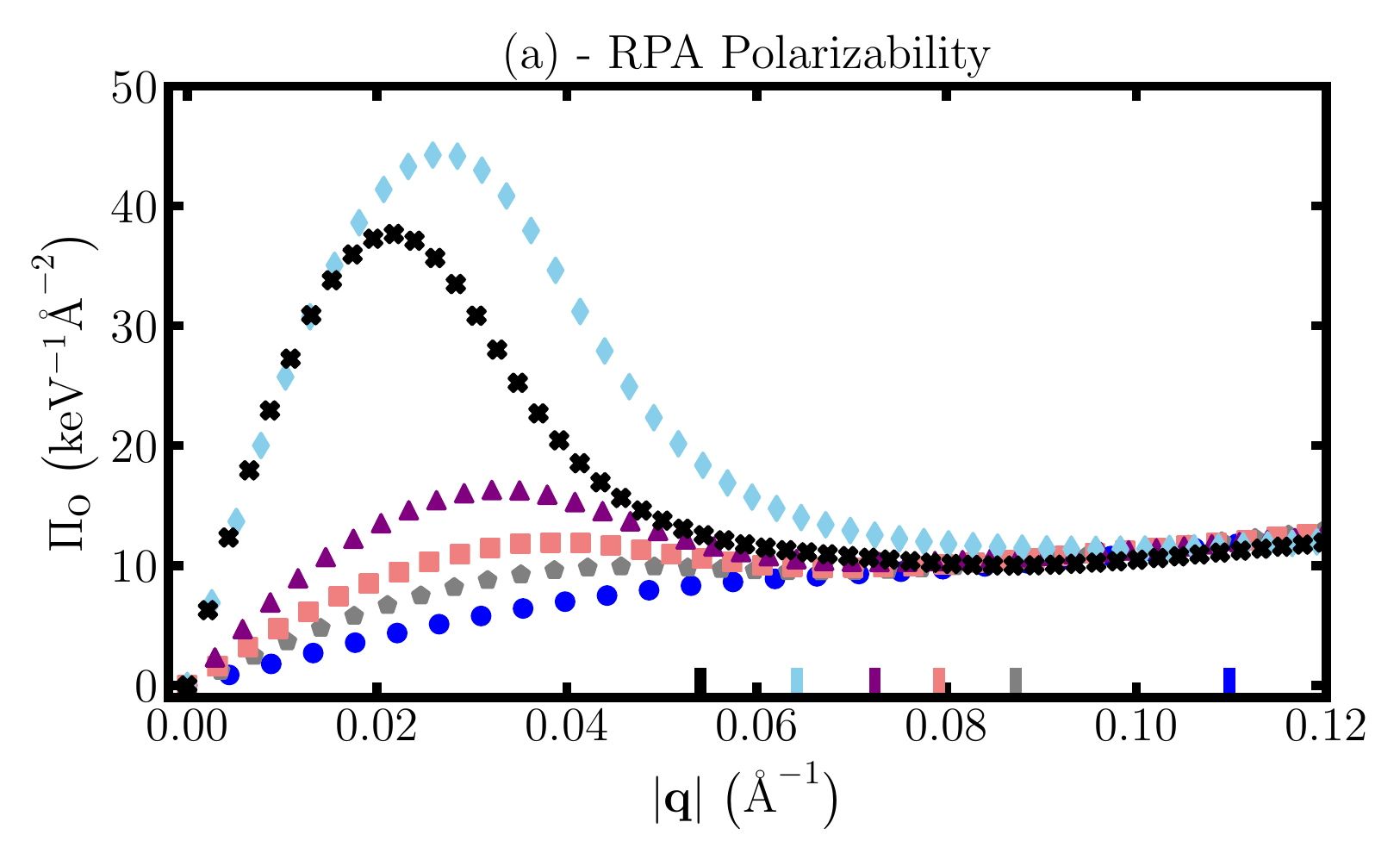}
\end{subfigure}
\begin{subfigure}{0.49\textwidth}
  \centering
  \includegraphics[width=1\linewidth]{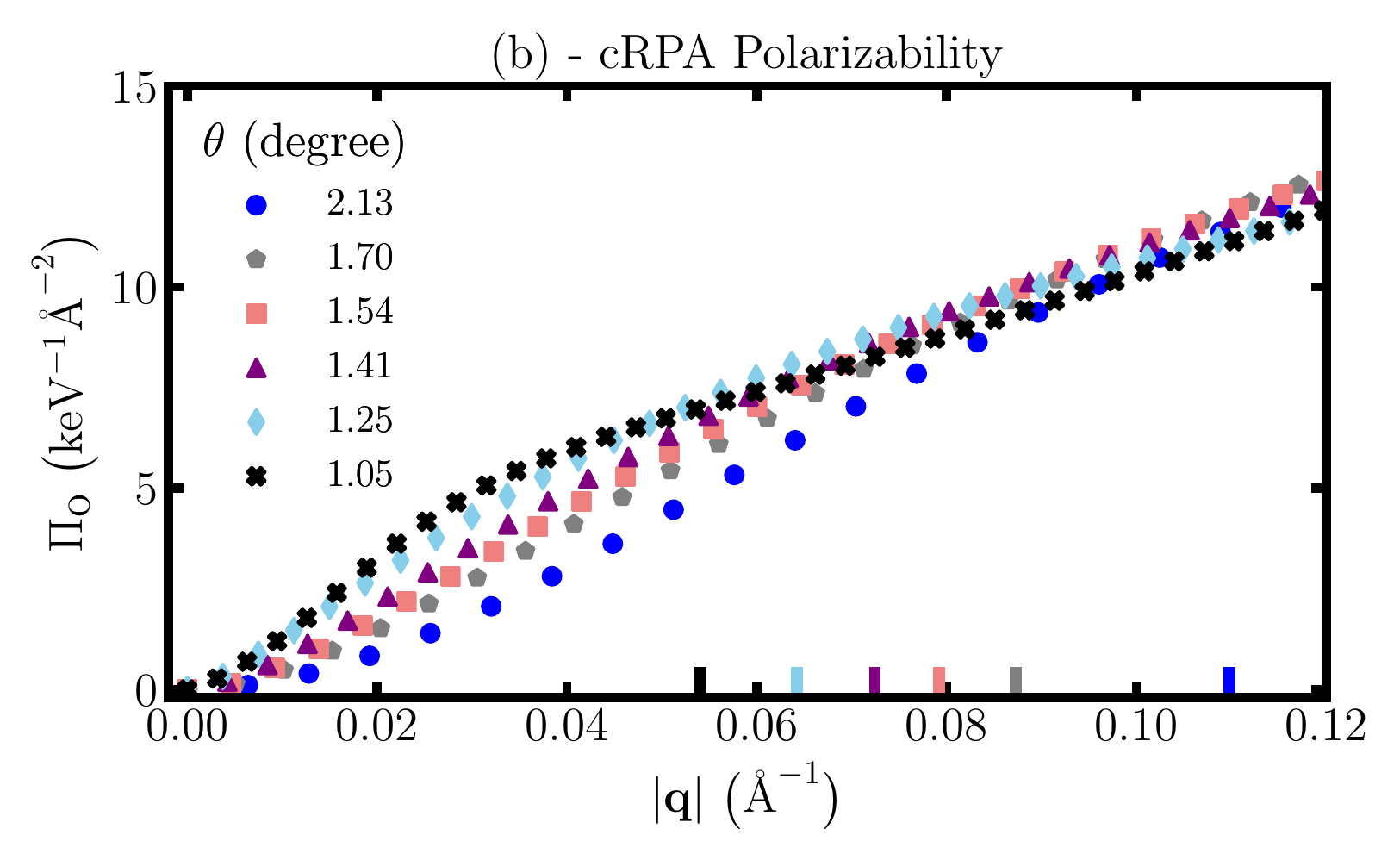}
\end{subfigure}
\\
\begin{subfigure}{0.49\textwidth}
  \centering
  \includegraphics[width=1\linewidth]{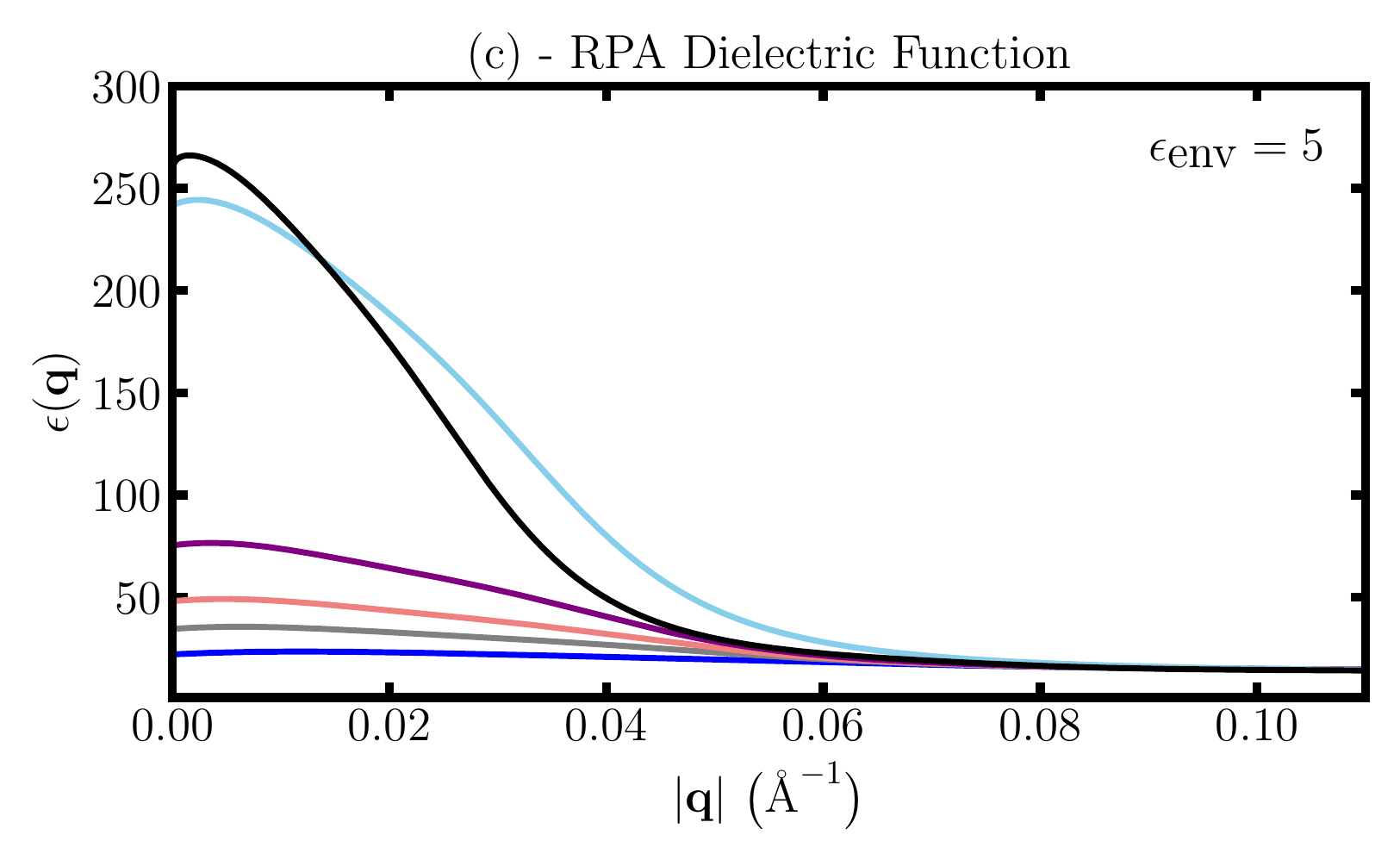}
\end{subfigure}
\begin{subfigure}{0.49\textwidth}
  \centering
  \includegraphics[width=1\linewidth]{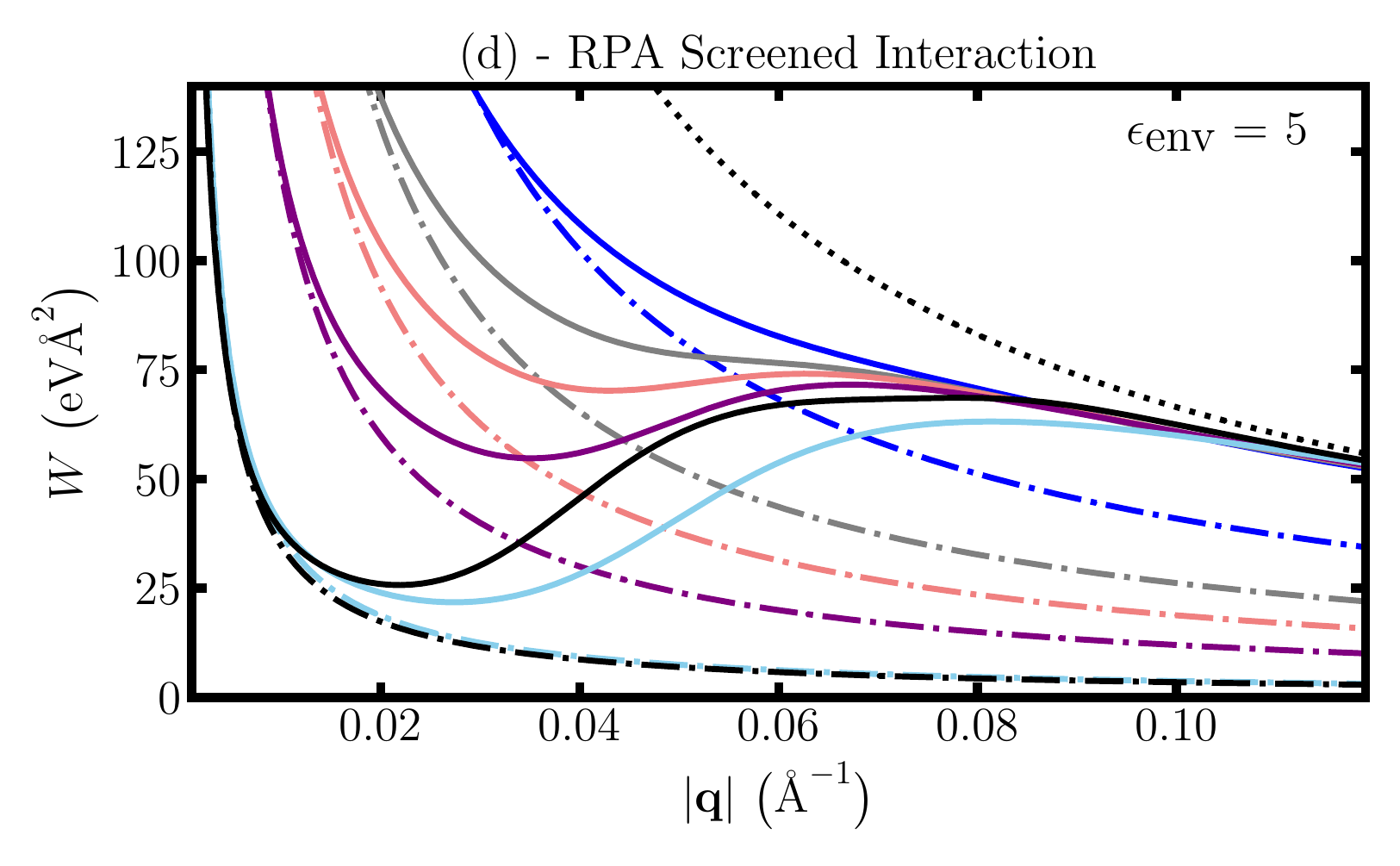}
\end{subfigure}
\\
\begin{subfigure}{0.49\textwidth}
  \centering
  \includegraphics[width=1\linewidth]{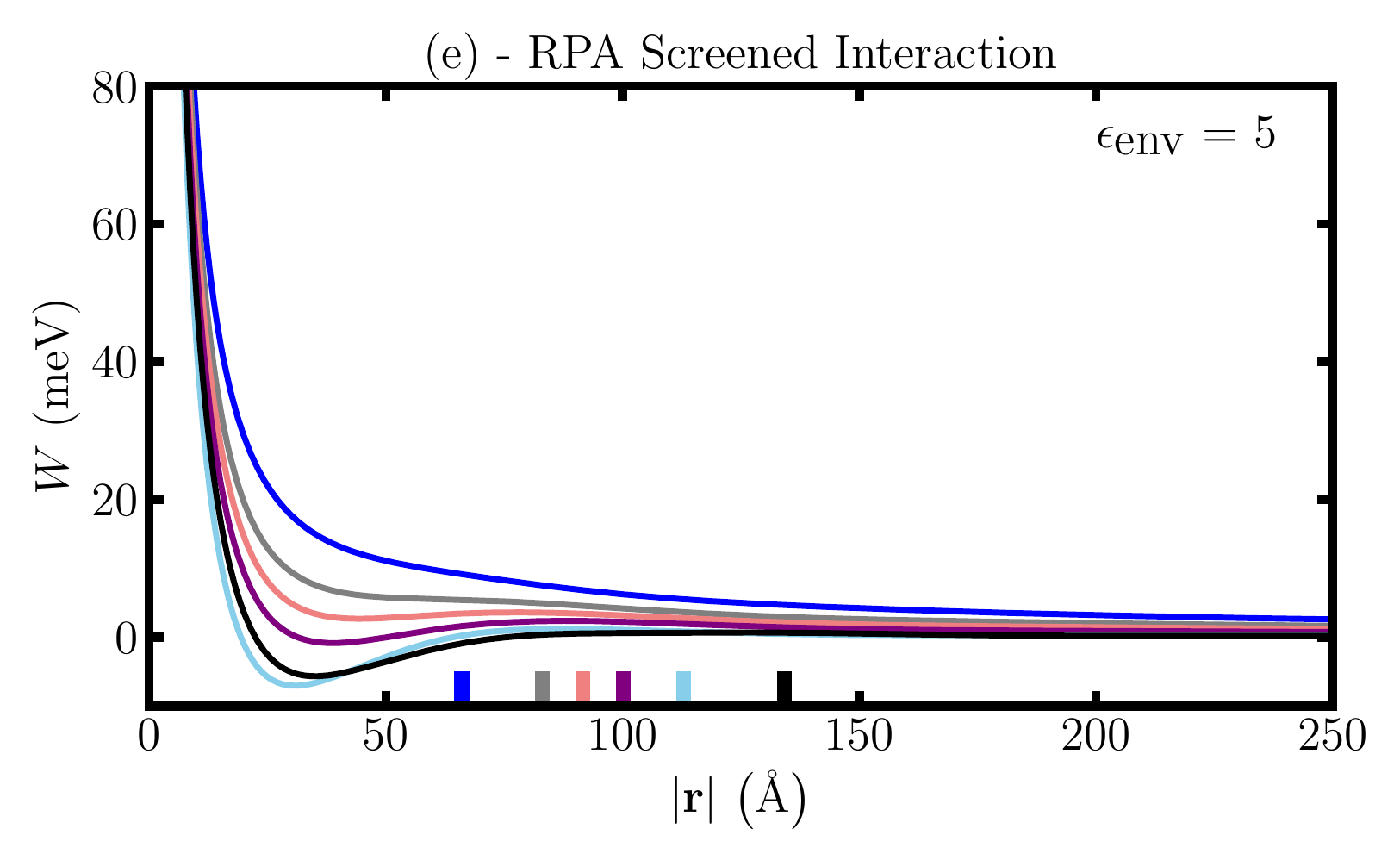}
\end{subfigure}
\begin{subfigure}{0.49\textwidth}
  \centering
  \includegraphics[width=1\linewidth]{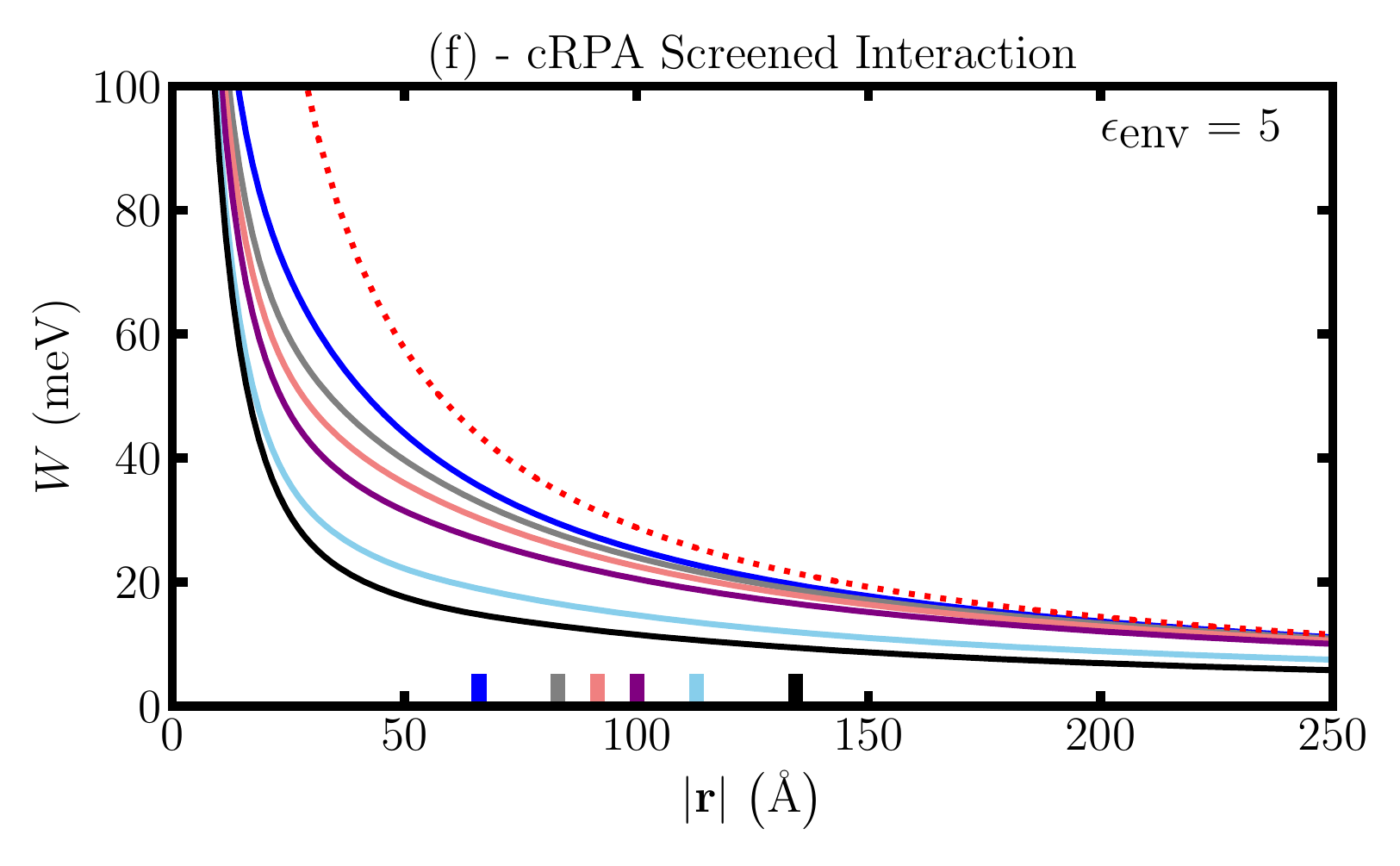}
\end{subfigure}
\caption{(a) and (b): RPA and cRPA polarizability of tBLG as a function of momentum transfer for several twist angles near the magic angle. Vertical stubs indicate the magnitude of first moir\'e reciprocal lattice vectors for each twist angle. (c) - RPA dielectric function of tBLG encapsulated in hBN as a function of wavevector. (d) - RPA screened interaction of tBLG encapsulated in hBN in momentum space (solid lines). Dash-dotted lines denote the long wavelength limit and the dotted line denotes the screened interaction divided by a dielectric constant with contributions from non-interacting graphene bilayers and hBN. (e) and (f): RPA and cRPA screened interaction in real space for tBLG encapsulated by hBN (solid lines). The red dash-dotted line indicates bare the Coulomb interaction. Vertical stubs denote moir\'e lattice constant for each twist angle.}
\label{POL_W}
\end{figure*}

\section{Results and Discussion}

\subsection{Polarizability and Screened Interaction}

Figure~\ref{POL_W}(a) shows the RPA polarizability of tBLG as function of crystal momentum for several twist angles in the vicinity of the magic angle. For these twist angles, we find that $\Pi^{\textrm{RPA}}_0 \propto q/v_{\text{F}}$ at small wavevectors as expected from the linear dispersion of the flat bands near K and K$^\prime$. The slope of $\Pi^{\textrm{RPA}}_0$ at small $\mathbf{q}$ depends sensitively on twist angle~\cite{PTBLG} because of the strong renormalization of the Fermi velocity, $v_{\text{F}}(\theta)$, which approaches zero at the magic angle~\cite{MBTBLG}. At wavevectors larger than the second reciprocal lattice vector of the moir\'e lattice, $\Pi^{\textrm{RPA}}_0$ of tBLG is very similar to that of decoupled graphene sheets~\cite{CCRPA}. In particular, it is linear in wave vector with a slope that is determined by the \textit{unrenormalized} Fermi velocity of graphene~\cite{CCRPA}. 

Cutting out transitions between flat bands from the RPA yields the cRPA polarizability, which is displayed in Fig.~\ref{POL_W}(b). The cRPA polarizability is highly isotropic and quadratic in $|\textbf{q}|$ for small $\textbf{q}$. This is characteristic of 2D semiconductors, such as molybdenum disulfide, and a consequence of the finite energy gap for transitions in the cRPA. The polarizability at small wavevectors increases with decreasing twist angle because the energy gap between the non-flat bands decreases, as seen in Fig.~\ref{BSBW}(b).

\begin{figure*}[t!]
\begin{subfigure}{0.49\textwidth}
  \centering
  \includegraphics[width=1\linewidth]{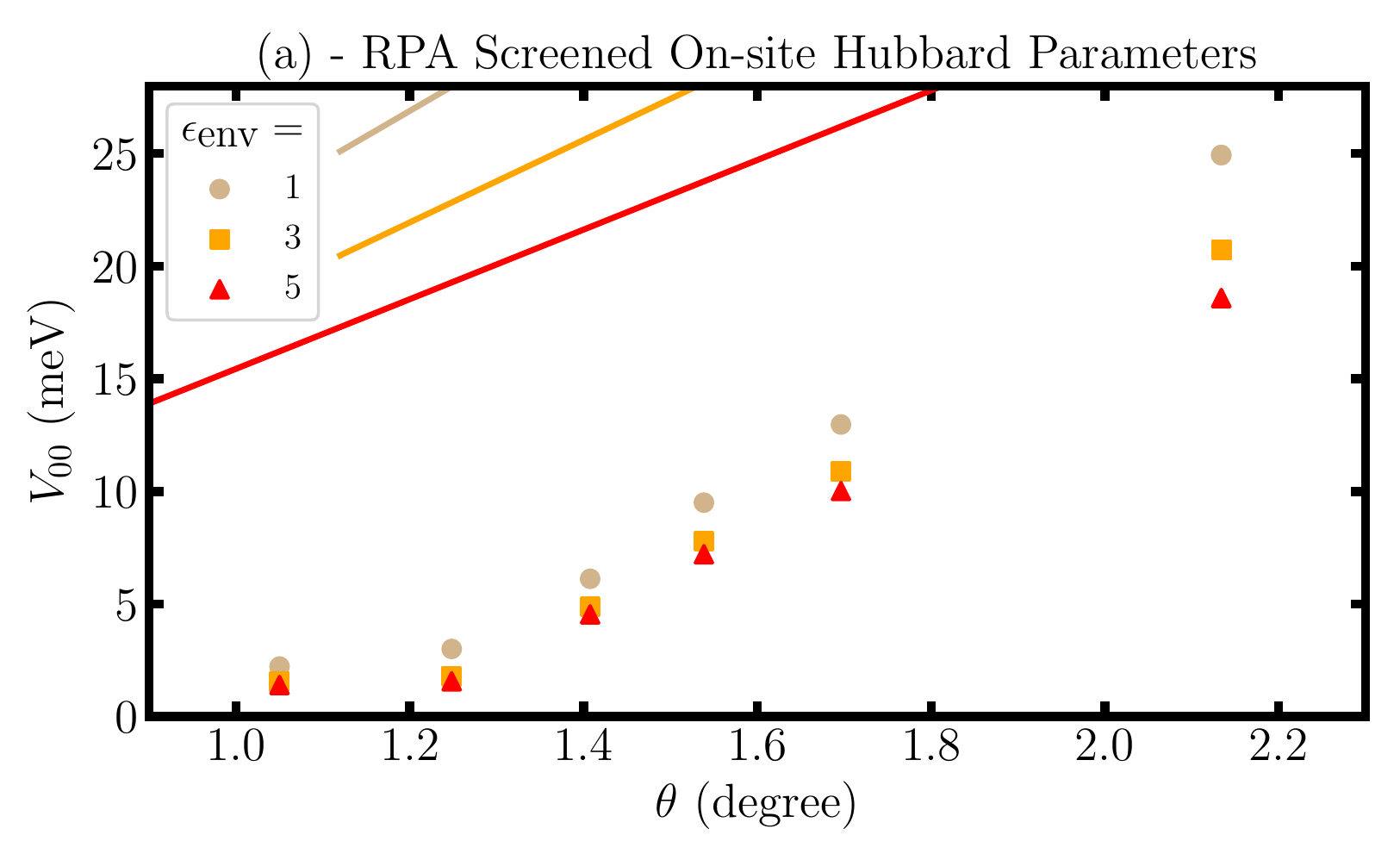}
\end{subfigure}
\begin{subfigure}{0.49\textwidth}
  \centering
  \includegraphics[width=1\linewidth]{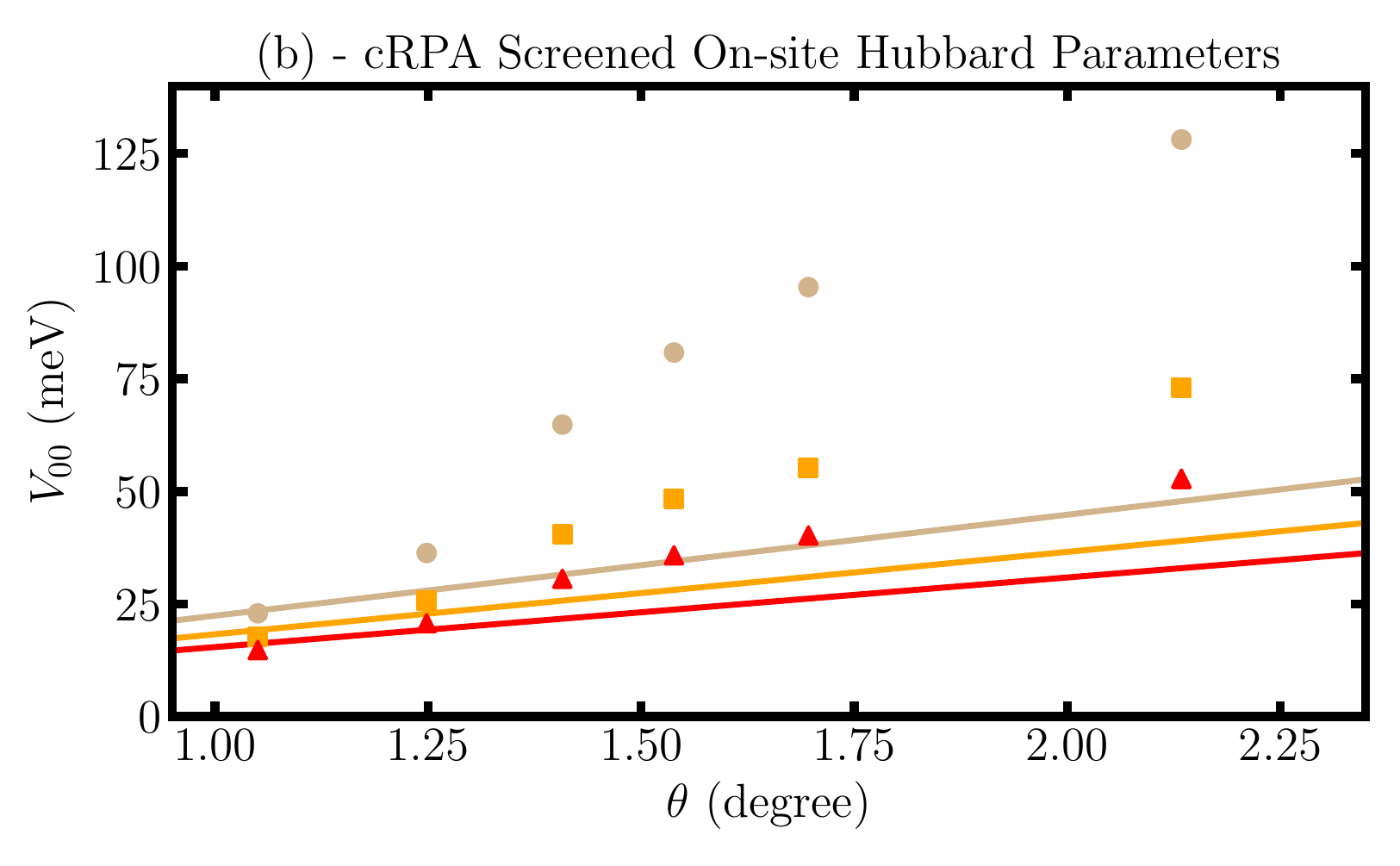}
\end{subfigure}
\\
\begin{subfigure}{0.49\textwidth}
  \centering
  \includegraphics[width=1\linewidth]{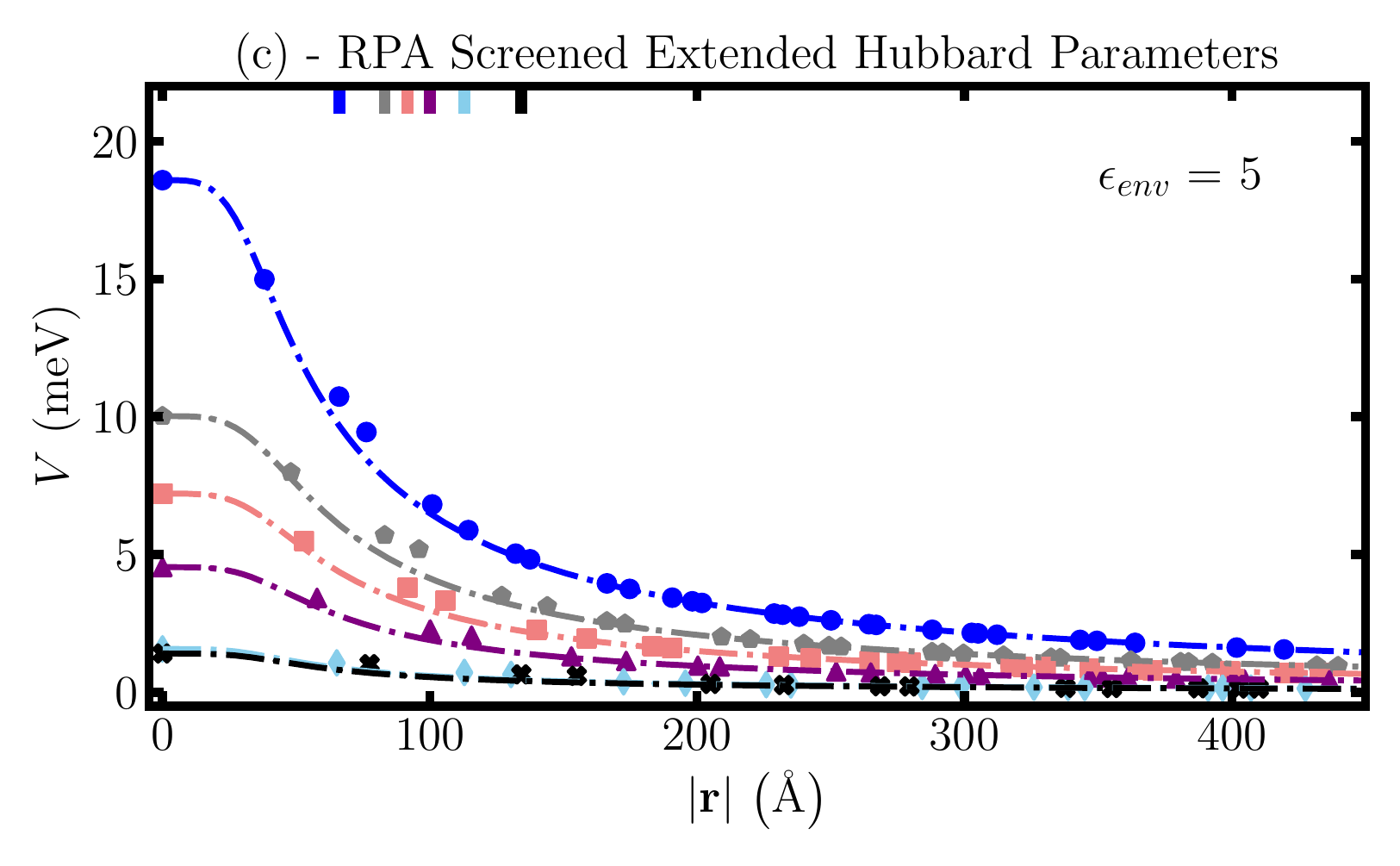}
\end{subfigure}
\begin{subfigure}{0.49\textwidth}
  \centering
  \includegraphics[width=1\linewidth]{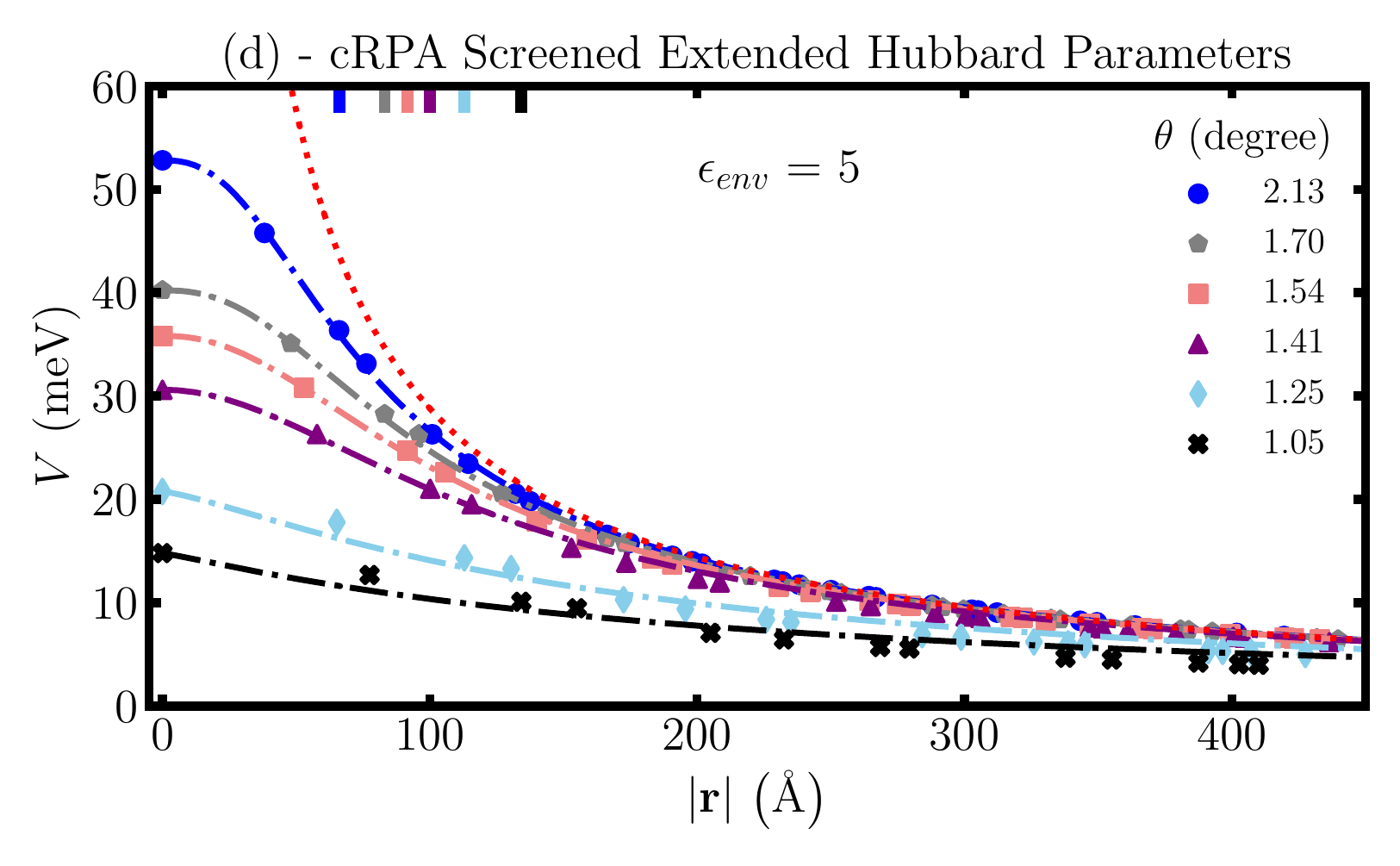}
\end{subfigure}
\caption{(a) and (b): RPA and cRPA screened on-site Hubbard parameters (symbols) as a function of twist angle for several environmental dielectric constants. Solid lines denote fits to bare on-site Hubbard parameters from Ref.~\citenum{PHD_1} divided by a dielectric constant with contributions from environmental screening plus decoupled graphene bilayers. (c) and (d): RPA and cRPA screened extended Hubbard parameters (symbols) as a function of Wannier function separation for several twist angles. Dash-dotted lines denote generalized Ohno potential fits; the dotted red line denotes the hBN screened Coulomb potential and vertical stubs denote the size of the moir\'e lattice vector.}
\label{HUB}
\end{figure*}

Table~\ref{DC} shows the twist-angle dependent value of the screening parameter $\alpha(\theta)$, obtained from fitting the quadratic polarizability at small $\textbf{q}$, that enters the widely used Keldysh model for the dielectric function of 2D semiconductors, $\epsilon(q) = 1 + \alpha q$~\cite{KEL}. At small twist angles, we find a dramatic increase of the screening parameter reaching values of more than 1000~$\textrm{\AA}$. This indicates that the Coulomb interaction is screened already for very small wavevectors (those larger than $1/\alpha$). At crystal momenta larger than the first moir\'e reciprocal lattice vector, $\Pi^{\textrm{cRPA}}_0$ also becomes linear in $|\mathbf{q}|$ and very similar to the polarizability of uncoupled graphene bilayers~\cite{CCRPA}.

\begin{table}[htp]
\begin{center}
\begin{tabular}{lcccr}
\hline\hline
$\theta$ / degree & $\quad\quad$ & $\epsilon$ (RPA) & $\quad\quad$ & $\alpha$ / $\textrm{\AA}$ (cRPA)\\
\hline 
2.13 & & 18.1 & & 155.4 \\
1.70 & & 30.6 & & 255.2 \\
1.54 & & 44.1 & & 327.8 \\
1.41 & & 71.4 & & 430.4 \\
1.25 & & 237.5 & & 889.3 \\
1.05 & & 256.3 & & 1292.0\\
\hline\hline
\end{tabular}
\end{center}
\caption{Long wavelength dielectric constants of tBLG from linear fits to Fig.~\ref{POL}(a) (with $\epsilon_{\text{env}}$ is taken to be 1) and Keldysh parameters from quadratic fits to Fig.~\ref{POL}(b) as a function of twist angle.}
\label{DC}
\end{table}

The RPA dielectric functions are displayed in Fig.~\ref{POL_W}(c). For angles close to the magic angle, the large linear slope of the polarizability at small wave vectors gives rise to a large dielectric constant. At larger wave vectors, the reduced slope of $\Pi_0^\text{RPA}$ results in a significantly smaller dielectric constant. The crossover between these two regimes of approximately constant dielectric functions occurs on the scale of the first two reciprocal moir\'e lattice vectors. Table~\ref{DC} shows the resulting long-wavelength dielectric constants $\epsilon(\theta)$ of tBLG in air ($\epsilon_{\textrm{env}}=1$). All angles exhibit enhanced dielectric constants relative to decoupled graphene bilayers ($\epsilon_\textrm{ni}=8.86$~\cite{PTBLG,EPG}). Near the magic angle, the dielectric constant of tBLG increases dramatically and reaches values larger than 250 - a factor of 20 larger than decoupled graphene bilayers.

The RPA screened interaction in momentum space of tBLG encapsulated by a dielectric substrate ($\epsilon_{\textrm{env}}=5$) is shown in Fig.~\ref{POL_W}(d). The effective interaction crosses over from a strongly screened small wave vector regime to a less strongly screened large wave vector regime. As a consequence of this crossover, the interaction exhibits a well-like feature for twist angles near the magic angle. Fig.~\ref{POL_W}(e) shows the interaction after Fourier transformation to real space. At several twist angles near the magic angle, the screened interaction in real space exhibits an \textit{attractive region}. Specifically, the minimum of the screened interaction occurs near 40~\AA~with a well depth of up to $\sim 10$~meV. The origin and consequences of this attractive region are discussed below (see also Appendix~\ref{AIS}). At larger separations, the screened interaction decays as $1/(\epsilon(\theta)|\mathbf{r}|)$, i.e., it is repulsive and significantly weaker than the screened interaction of uncoupled graphene bilayers. At small separations, the screened interaction of tBLG is similar to that of uncoupled bilayers. 

Fig.~\ref{POL_W}(f) shows the real-space cRPA screened interaction. At small distances, the interaction is similar to that of uncoupled graphene bilayers, while at large distances it proportional to $1/(\epsilon_{\textrm{env}}|\mathbf{r}|)$, i.e., the bare interaction screened by the dielectric constant of the environment (red dotted line). The distance at which the crossover between these two regimes occurs is determined by the twist-angle dependent Keldysh parameter $\alpha(\theta)$, see discussion above.

Upon doping tBLG, intra-band transitions will occur (in addition to the inter-band transition studied in the current manuscript) in the RPA. These transitions will give rise to metallic screening – similar to the case of graphene where an analytical expression of the RPA dielectric function can be obtained~\cite{DPGFD}. Naively, one could simply adapt this expression to the case of twisted bilayer graphene by modifying the degeneracy factor (to take into account that there are two layers) and the renormalization of the Fermi velocity. Such a treatment, however, would not capture the attractive regions in the screened interaction which arise from rapid changes in the Fermi velocity. To understand what happens to these regions when the system is doped, we have analysed a model dielectric function in Appendix~\ref{AIS}. We found that the attractive regions should persist when electrons or holes are added suggesting that they could indeed play an important role for the correlated insulator states or superconductivity. For the cRPA screened interaction, we do not expect significant changes upon doping since the doping only affects the flat bands and there is a significant gap between those and all other bands.

\subsection{Hubbard Parameters}

The increased internal screening combined with the emergence of attractive regions in the RPA interaction leads to a significant reduction of the on-site and extended interaction parameters~\cite{PHD_1}. Fig.~\ref{HUB}(a) shows the screened on-site Hubbard parameters, $V_{00}$, as function of twist angle for different values of the environmental dielectric constant $\epsilon_{\textrm{env}}$, and compares them to the linear fits to the on-site Hubbard parameters calculated with a Coulomb potential screened by a dielectric constant with contributions from the environment and uncoupled bilayers (solid lines). In contrast to the case of uncoupled bilayers, the RPA on-site Hubbard parameters are relatively small near the magic angle (only a few meV instead of tens of meV), and they are a non-linear functions of twist angle~\cite{PHD_1}.

Similarly, the extended Hubbard parameters for tBLG, shown in Fig.~\ref{HUB}(c), are strongly reduced near the magic angle compared to uncoupled graphene bilayers~\cite{PHD_1}. The calculated Hubbard parameters are well described by an analytical Ohno-like expression $V(r) = V_{00}/\sqrt[4]{1 + (V_{00}/W_{\textrm{RPA}}(r))^{4}}$, where $W_{\textrm{RPA}}(r)$ denotes the screened RPA interaction in the long wavelength limit and $r$ is the separation between Wannier function centers. 

The Hubbard parameters obtained from the cRPA interaction are shown in Figs.~\ref{HUB}(b) and (d). The on-site Hubbard parameters from the cRPA interaction, as displayed in Fig.~\ref{HUB}(b), are approximately one order of magnitude larger than the RPA values and display a non-linear dependence on twist angle. In contrast, the simplified screening model exhibits a linear dependence~\cite{PHD_1}. The extended cRPA Hubbard parameters, shown in Fig.~\ref{HUB}(d), approach the bare Coulomb interaction divided by the environmental dielectric constant (red dotted line) at large Wannier function separations, and are well-described by the analytical Ohno-like model $V(r) = V_{00}/\sqrt[n]{1 + (V_{00}/W_{\textrm{env}}(r))^{n}}$, where the exponent $n$ is fitted for each twist angle separately and $W_{\textrm{env}}(r) \propto 1/(\epsilon_\text{env}r)$ (see Appendix~\ref{AWF} for details).

\begin{figure}[h!]
\centering
\includegraphics[width=1\linewidth]{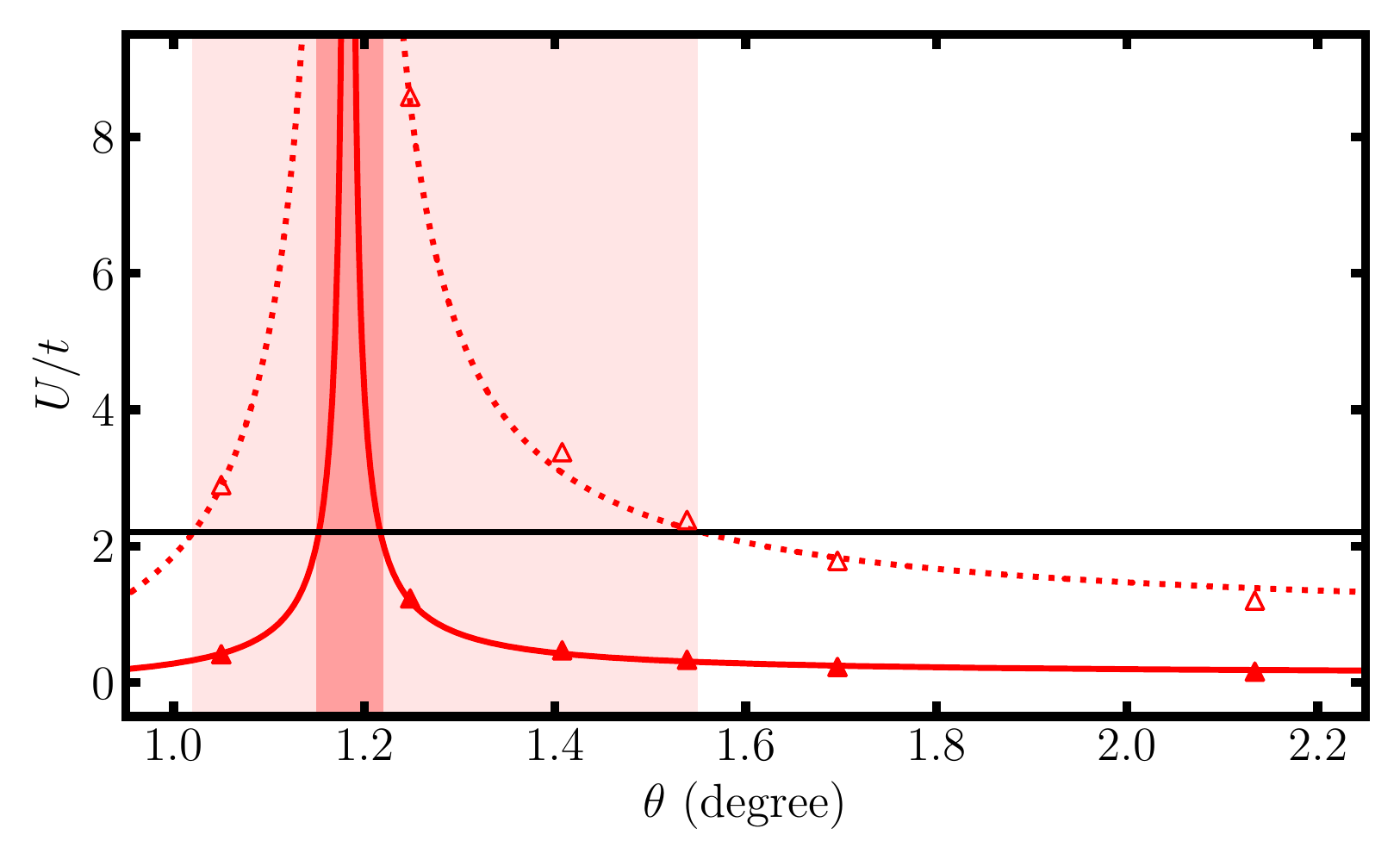}
\caption{$U/t$ of tBLG as function of twist angle (open symbols) with $U$ being the cRPA on-site Hubbard parameter and $t$ denoting the hopping parameter. Closed symbols indicate $U^*/t$ (with $U^*$ denoting the difference between the on-site and the nearest-neighbour cRPA Hubbard parameters) which is a better measure of electron correlations in systems with long-range interactions.}
\label{UT}
\end{figure}

\subsection{Discussion}

In this section, we discuss the origin of the attractive regions in the RPA screened interaction and the consequences for the phase diagram of tBLG. 

Attractive regions in the screened interaction are also found in the two-dimensional and three-dimensional electron gas where they are a consequence of Friedel oscillations~\cite{DPGFD}. These real-space oscillations of the induced charge density are caused by the discontinuity of the Bloch state occupancy at the Fermi level in $k$-space. Importantly, undoped graphene and undoped tBLG (at the twist angles we study) do not exhibit Friedel oscillations because the density of states at the Fermi level vanishes (recall, however, that we do find metallic band structures of undoped tBLG for certain twist angles, and tBLG away from charge neutrality might be expected to exhibit such oscillations). Instead, the attractive regions in tBLG have a different origin: they are caused by the abrupt change of the band velocity as a function of the band energy which gives rise to the peaks in the RPA polarizability, see Fig.~\ref{POL_W}(a). At small wave vectors, the RPA polarizability exhibits a large slope as a consequence of the strongly renormalized Fermi velocity of the flat bands~\cite{PTBLG}. At larger wave vectors, the slope of $\Pi^{RPA}_0$ reflects the unrenormalized Fermi velocity of uncoupled graphene sheets~\cite{CCRPA}. Fourier transformation of the resulting screened interaction to real space then results in oscillatory behaviour and attractive regions (see Appendix~\ref{AIS} for further details).

The screened interaction influences many properties of tBLG. For example, it determines the interaction of charged defects with the electrons in tBLG which can be studied with transport measurements and scanning tunnelling spectroscopy and microscopy techniques~\cite{FC,MMCI}. Moreover, photo-excited electron-hole pairs interact via the screened interaction giving rise to excitonic effects in the optical properties of tBLG~\cite{LPG}. Attractive regions in the screened interaction can also induce electronic phase transitions. It is well known that Friedel oscillations in 2D and 3D electron gases can give rise to Cooper pairing and superconductivity via the Kohn-Luttinger mechanism~\cite{KL,NMS,CHUB}. Similarly, superconductivity due to polarization induced electron pairing has been suggested to occur in long organic molecules with polarizable side chains~\cite{OS,EAMCR,ESQT}. Finally, electrons can reduce their potential energy by localizing in the attractive regions of the screened interaction leading to the formation of charge density waves. The resulting energy gaps could explain the recently observed correlated insulator behaviour in undoped tBLG~\cite{SOM}. Our calculations demonstrate that internal screening strongly reduces the on-site Hubbard parameter, see Fig.~\ref{HUB}(c). For graphene, Jung and MacDonald have shown that this favors the formation of charge density waves~\cite{ENLE}. 

Spin density waves have also been suggested as candidates for the correlated insulator states~\cite{SSPDG,SCDID,KVB,ECM,TJMODEL,US,CSD}. These phases are expected to occur when the ratio of the on-site Hubbard parameter $V_{00}$ (commonly denoted as $U$) and the hopping integral $t$ is large. Based on Quantum Monte calculations, Scalettar \textit{et al.}~\cite{PS} suggested that undoped tBLG undergoes a transition to a spin density wave at a $U/t$ value of about two. This agrees well with the critical value of $U/t=2.2$ obtained for Bernal stacked bilayers~\cite{ABS}. Here, we use a critical value of $U/t=2.2$, but stress that our qualitative conclusions do not depend on the precise choice for this value. As discussed above, the on-site Hubbard parameter that enters a downfolded Hamiltonian for the flat-band electrons should be screened by all transitions except those between flat bands. Fig.~\ref{UT} shows the ratio of $U$ calculated within the cRPA and the hopping parameter (approximated as $\Delta$/6\cite{PHD_1}) as function of the twist angle. $U/t$ exceeds the critical value of 2.2 in a significant twist-angle range ($\theta=1.02\degree$ to $\theta=1.52\degree$ corresponding to the light shaded region in the figure). It is well known, however, that long-ranged interactions reduce electron correlations. This effect can be approximately incorporated by replacing $V^\text{cRPA}_{00}$ by $U^*=V^\text{cRPA}_{00}-V^\text{cRPA}_{01}$, i.e., the difference between the on-site and nearest neighbour Hubbard parameters~\cite{OHP,PHD_1}. The resulting measure for the strength of electron correlations $U^*/t$ exceeds the critical value only in a very narrow range of twist angles (dark shaded region in the figure) indicating that spin density wave states can only be found in a narrow twist-angle window~\cite{PHD_1}.

\section{Conclusions}
We have calculated the screened interaction and extended Hubbard parameters in undoped tBLG for several twist angles in the vicinity of the magic angle using the random phase approximation (RPA) and also the constraint random phase approximation (cRPA). Near the magic angle, the flattening of the bands drastically increases the RPA dielectric constant of tBLG and also the cRPA Keldysh screening parameter. As a consequence, the extended Hubbard parameters depend sensitively on the twist angle and the on-site Hubbard is no longer a linear function of the twist angle. The abrupt change in the band velocity as function of the band energy gives rise to \textit{attractive regions} in the RPA screened interaction in real space which could induce novel effects in the optical and transport properties of tBLG. Moreover, the effective attraction between electrons can give rise to the formation of charge density waves and Cooper pairs, and thus be intimately connected to the correlated insulator states and superconducting phases that have been observed experimentally. These effects are not captured by Hartree-Fock calculations which employ a constant twist-angle independent dielectric function. 

\section{Acknowledgements}

We thank V. Vitale, D. Kennes, A. MacDonald, M. Scharnke, J. Knolle and A. Benyamini for helpful discussions. This work was supported through a studentship in the Centre for Doctoral Training on Theory and Simulation of Materials at Imperial College London funded by the EPSRC (EP/L015579/1). We acknowledge funding from EPSRC grant EP/S025324/1 and the Thomas Young Centre under grant number TYC-101.

\appendix

\section{Tight-Binding}
\label{ATB}

\renewcommand{\thefigure}{A\arabic{figure}}
\setcounter{figure}{0} 

\subsection{Moir\'e Structure}

We utilise an atomistic tight-binding model to calculate the electronic structure of twisted bilayer graphene (tBLG). This method requires finite unit cells associated with commensurate twist angles~\cite{LDE,NSCS}. Here we generate moir\'e unit cells by rotating the top graphene sheet of an AA stacked bilayer graphene around an axis perpendicular to the sheets that intersects a carbon atom in each layer. The resulting structure has D$_{3}$ symmetry. To generate a twist angle with a commensurate unit cell, an atom in the top, rotated graphene layer must reside exactly above another atom in the bottom, unrotated graphene sheet. The resulting commensurate lattice vectors of the moir\'e unit cell are given by $\textbf{R}_{1} = n\textbf{a}_{1} + m \textbf{a}_{2}$ and $\textbf{R}_{2} = -m\textbf{a}_{1} + (n + m) \textbf{a}_{2}$, where $n$ and $m$ are integers and $\textbf{a}_{1} = (\sqrt{3}/2, -1/2)a_{0}$ and $\textbf{a}_{2} = (\sqrt{3}/2, 1/2)a_{0}$ denote the lattice vectors of graphene (with $a_{0} = 2.46$ $\textrm{\AA}$)~\cite{LDE,NSCS}.
The corresponding twist angle, $\theta$, is given by
\begin{equation}
\cos\theta = \dfrac{n^{2} + 4nm + m^{2}}{2(n^{2} + nm + m^{2})}.
\end{equation}

Significant out-of-plane lattice relaxations occur in tBLG at small twist angles~\cite{AC,LSDFT,PDTBLG,SETLA,STBBG,EDS,EPC,LREBM,MLWO,CTBS}; in-plane relaxations also occur, but the magnitude of the relaxation is smaller. Here, we only take out-of-plane relaxations into account. Specifically, we employ the following expression from Ref.~\citenum{MLWO} for the out-of-plane atomic corrugation of carbon atoms at position $\textbf{r}$, 
\begin{equation}
z(\textbf{r}) = d_{0} + 2d_{1}\sum_{i = 1,2,3}\cos(\textbf{b}_{i}\cdot\textbf{r}).
\end{equation}

\noindent Here, $\textbf{b}_{1}$ and $\textbf{b}_2$ denote the primitive moir\'e reciprocal lattice vectors of tBLG, and $\textbf{b}_{3}=\textbf{b}_1+\textbf{b}_2$. Also $d_{0} = (d_{AA} + 2d_{AB})/3$ and $d_{1} = (d_{AA} - d_{AB})/9$ with $d_{AB} = 3.35$ $\textrm{\AA}$ and $d_{AA} = 3.60$ $\textrm{\AA}$, respectively, being the interlayer separations of AB and AA stacked bilayer graphene~\cite{MLWO}. 

\subsection{Hamiltonian and Band Structure}

For this atomic structure of tBLG, we solve the atomistic tight-binding Hamiltonian~\cite{FC,PHD_1}
\begin{equation}
\mathcal{\hat{H}}_{0} = \sum_{i}\epsilon_{i}\hat{c}^{\dagger}_{i}\hat{c}_{i} + \sum_{i,j}(t(\textbf{r}_{i} - \textbf{r}_{j})\hat{c}^{\dagger}_{j}\hat{c}_{i} + \text{H.c.}),
\end{equation}

\noindent where $\epsilon_{i}$ is the on-site energy of the $p_{z}$-orbital on atom $i$ (which is set to zero in our calculations), and $\hat{c}^{\dagger}_{i}$ and $\hat{c}_{i}$ denote creation and annihilation operators of electrons in the $p_z$-orbital on atom $i$, respectively. Also, $t(\textbf{r}_{i} - \textbf{r}_{j})$ denotes the hopping parameter between atoms $i$ and $j$~\cite{LDE,NSCS}. To calculate the hopping parameters, we employ the Slater-Koster rules~\cite{SK,LDE}, i.e.,
\begin{equation}
t(\textbf{r}) = V_{pp\sigma}(\textbf{r})n^{2} + V_{pp\pi}(\textbf{r})(1 - n^{2}),
\end{equation}

\noindent where $n = \textbf{r}\cdot\textbf{e}_{z}/|\textbf{r}|$. The intra-layer hopping is described by the $\pi$-bonding character of $p_{z}$-orbitals
\begin{equation}
V_{pp\pi} = V_{pp\pi}^{0}e^{q_{\pi}(1 - |\textbf{r}|/a)},
\end{equation}

\noindent where $V_{pp\pi}^{0} = -2.7$ eV is the nearest-neighbour hopping parameter in graphene (for an equilibrium bond length of $a = 1.42$ $\textrm{\AA}$~\cite{EPG}) and $q_{\pi} = 3.14$ describes the decay of the hopping as function of distance~\cite{LDE,NSCS}. After third-nearest neighbours, we set the hopping to zero. The inter-layer coupling has contributions from both $\pi$ and $\sigma$ interactions of $p_{z}$-orbitals, with the latter given by 
\begin{equation}
V_{pp\sigma} = V_{pp\sigma}^{0}e^{q_{\sigma}(1 - |\textbf{r}|/d_{AB})},
\end{equation}

\noindent where $V_{pp\sigma}^{0} = 0.48$ eV is the magnitude of the hopping in the AB/BA regions and  $q_{\sigma} = 7.43$ is the corresponding decay length of this hopping~\cite{LDE,NSCS}. After an in-plane distance corresponding to third-nearest neighbours, we set all inter-layer hopping parameters to zero. 

The Bloch eigenstates of the tight-binding Hamiltonian are
\begin{equation}
 \psi_{n\textbf{k}}(\textbf{r}) = \frac{1}{\sqrt{N}}\sum_{j\textbf{R}}c_{jn\textbf{k}}e^{i\textbf{k}\cdot \textbf{R}}\phi_{z}(\textbf{r} - \textbf{t$_{j}$} - \textbf{R}),
\end{equation}

\noindent where $\phi_z$ denotes the pseudo-hydrogenic wavefunction of the p$_z$-orbital, $\textbf{t$_{j}$}$ is the position of carbon atom $j$ in the unit cell, $N$ denotes the number of moir\'e unit cells in the crystal and $c_{jn\textbf{k}}$ are coefficients obtained from the diagonalization of the Hamiltonian. 

\begin{figure*}[t!]
\begin{center}
\includegraphics[width = 0.9\textwidth]{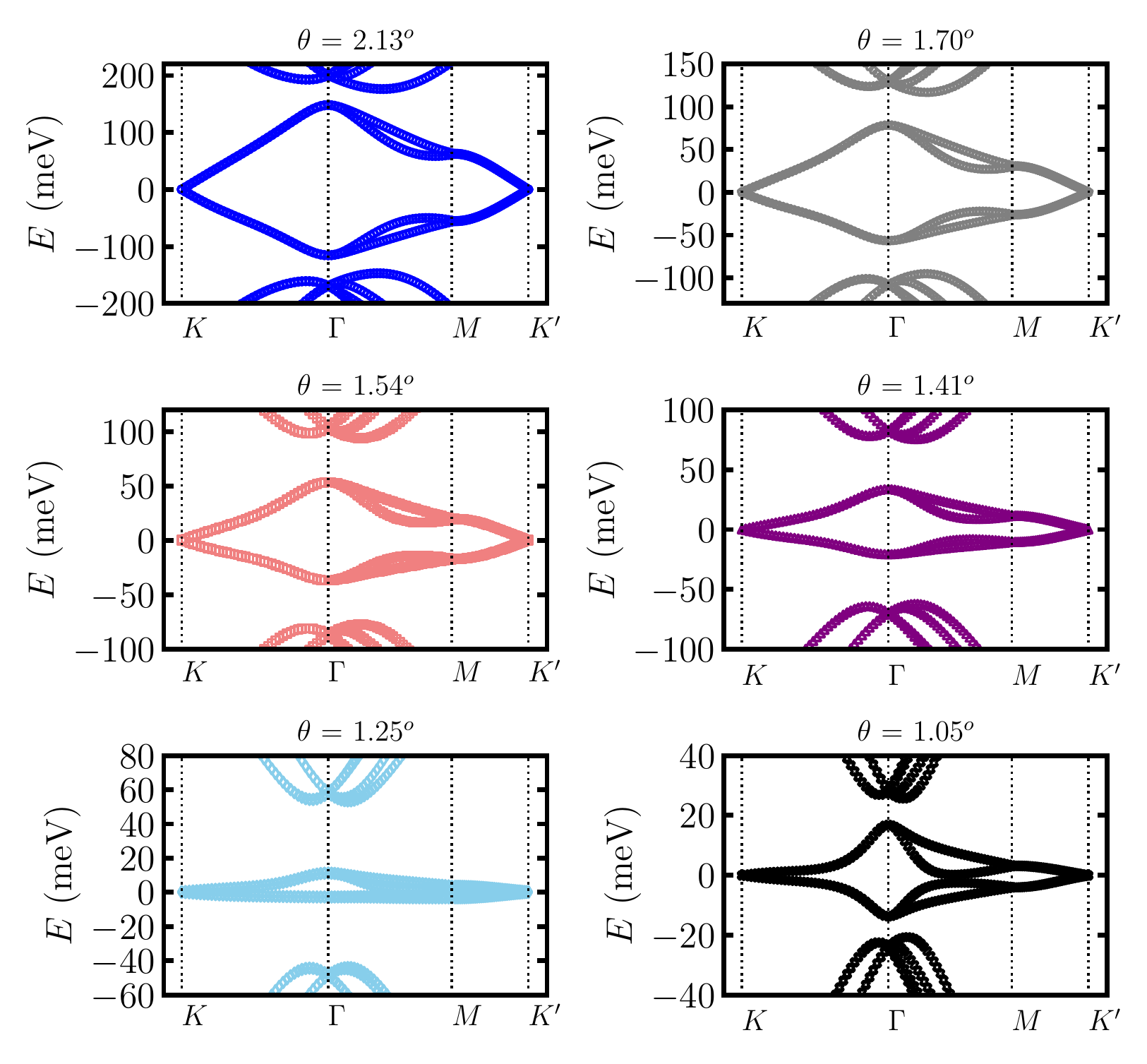}\\
\caption{Band structures of twisted bilayer graphene for the twist angles studied in the main text. Subfigure for the twist angle 1.05$\degree$ adapted with permission from Ref.~\citenum{PHD_1}. Copyrighted by the American Physical Society.}
\label{BS_FIG}
\end{center}
\end{figure*}

In Fig.~\ref{BS_FIG} the resulting band structures are shown for the twist angles under consideration (see main text). These band structures are in good agreement with others in the literature.

\subsection{Band Width and Band Gap Fitting}

Figure~\ref{UT} of the main text shows the ratio of on-site Hubbard parameters and the hopping parameter (given by the band width divided by six) to identify the twist angle ranges where spin density waves at charge neutrality should emerge. To obtain the continuous line, we fit the band width near the magic angle, $\theta^{*} = 1.18\degree$~\cite{PHD_1}, with the following equation
\begin{equation}
\Delta = \delta |\theta - \theta^{*}|.
\end{equation}

\noindent We find $\delta=0.27$ eV/degree. The band gap between non-flat bands at $\Gamma$ was fitted to a straight line.

\section{Internal Screening}
\label{AIS}

\renewcommand{\thefigure}{B\arabic{figure}}
\setcounter{figure}{0} 

\subsection{Polarizability Calculation}

To calculate the polarizability, as shown by Eq.~\eqref{POL} of the main text, a sum over all $k$-points in the first Brillouin zone and all transitions from occupied valence bands to unoccupied conduction bands must be performed. For this, the matrix elements, $\braket{\psi_{n^{\prime}\textbf{k}+\textbf{q}}|e^{i\textbf{q}\cdot\textbf{r}}|\psi_{n\textbf{k}}}$, must be determined. Inserting the tight-binding expression for the Bloch states and neglecting contributions from pairs of orbitals that do not sit on the same atom, we find
\begin{equation}
\braket{\psi_{n^{\prime}\textbf{k}+\textbf{q}}|e^{i\textbf{q}\cdot\textbf{r}}|\psi_{n\textbf{k}}} = \sum_{i}c^{*}_{n^{\prime}\textbf{k}+\textbf{q}i}c_{n\textbf{k}i}e^{i\textbf{q}\cdot\textbf{t$_{i}$ }} I(\textbf{q}),
\end{equation}

\noindent where the integral $I(\mathbf{q})$ is given by
\begin{equation}
I(\textbf{q}) = \int d\textbf{r}\phi^{*}_{z}(\textbf{r})e^{i\textbf{q}\textbf{r}}\phi_{z}(\textbf{r}) = \bigg[\dfrac{1}{1 + (|\textbf{q}|a_{0}/Z)^{2}}\bigg]^{3}.
\end{equation}

\noindent Here, $a_{0}$ is the Bohr radius and $Z$ is the effective charge of the carbon atom, taken to be 3.18~\cite{Shung}. Note that for the crystal momenta studied in this work, this integral can be safely set to 1. 

\subsection{Long-wavelength limit}

To parameterise the Keldysh model we used three different 7$\times$7 grids: one of these which contained the $\Gamma$ point, and two that were shifted by $0.05(\textbf{b}_1 + \textbf{b}_2)$ and $0.01(\textbf{b}_1 + \textbf{b}_2)$. By calculating transitions between these grids were we able to fit a quadratic curve in the long wavelength limit.

\subsection{Non-interaction dielectric constant}

The polarizability of non-interacting graphene bilayer is given by 
\begin{equation}
\Pi_{0}(\textbf{q}) = \dfrac{g_{s}g_{v}g_{l}|\textbf{q}|}{16\gamma},
\end{equation}
\noindent where $g_s$, $g_v$ and $g_l$ are the spin, valley and layer degeneracy, respectively, all of which are equal to 2, and $\gamma$ is the band parameter~\cite{PTBLG,SEISMG}, where $\gamma$ is related to the hopping parameter of graphene, $t_G = 2.7$~eV, and the bond length, $a = 1.42 ~\textrm{\AA}$, through $\gamma = 3t_{G}a/2$. Inserting these values into the equation for dielectric function yields
\begin{equation}
    \epsilon_{\textrm{ni}} = 1 + \dfrac{e}{6\epsilon_0 t_Ga} \approx 8.86.
\end{equation}
\noindent Note that the hopping parameter of graphene can vary, and this can yield different results for the dielectric constant~\cite{SEISMG,EPG}.

\subsection{Real-Space Screened Interaction}

Since the polarizability was found to be approximately isotropic, Eq.~\eqref{WIR} of the main text can be transformed to 
\begin{equation}
W(r) = \dfrac{e^{2}}{4\pi\epsilon_{0}}\int_{0}^{\infty} dq\dfrac{J_{0}(qr)}{\epsilon(q)},
\label{WJ}
\end{equation}

\noindent where $q$ and $r$ denote the magnitudes of the in-plane momentum and the in-plane distance, respectively, and $J_{0}$ is a Bessel function of the first kind with zeroth order. 

The calculated polarizabilities exhibit two regimes as function of crystal momentum: at large momenta (i.e. those larger than twice the length of the primitive reciprocal lattice vectors), tBLG responds similar to decoupled bilayer graphene; at small momenta, a significant enhancement in the response as a function of twist angle is observed. Therefore, the integral of Eq.~\eqref{WJ} can be separated into two parts,
\begin{equation}
\begin{split}
W(r) &= \dfrac{e^{2}}{4\pi\epsilon_{0}}\Bigg[\int_{0}^{2|\textbf{b}|} dq \dfrac{J_{0}(qr)}{\epsilon(q)} + \int_{2|\textbf{b}|}^{\infty} dq \dfrac{J_{0}(qr)}{\epsilon(q)}\Bigg] \\ &= W_{\text{s}}(r) + W_{\text{l}}(r),
\end{split}
\end{equation}

\noindent where $|\mathbf{b}|$ denotes the length of the primitive reciprocal lattice vectors. The first contribution, $W_{\text{s}}$, stems from the response at small wavevectors, which can be numerically integrated and readily converged. The second contribution, $W_{\text{l}}$, is the contribution from large momenta, which is essentially that of decoupled bilayer graphene. Since the dielectric function is a constant in the latter regime, $\epsilon(q \ge 2|\textbf{b}|) \approx \epsilon_{\mathrm{ni}}$, the integral can be transformed to 
\begin{equation}
W_{\text{l}} = \dfrac{e^{2}}{4\pi\epsilon_{0}\epsilon_{\text{ni}}r}\Bigg[1 -  \int_{0}^{2|\textbf{b}|r} dy J_{0}(y)\Bigg],
\end{equation}

\noindent which can also be readily evaluated. Larger cut-off values for separating the small and large momentum regimes were also used and found to not alter the result.

\begin{figure*}[t!]
\begin{subfigure}{0.49\textwidth}
  \centering
  \includegraphics[width=1\linewidth]{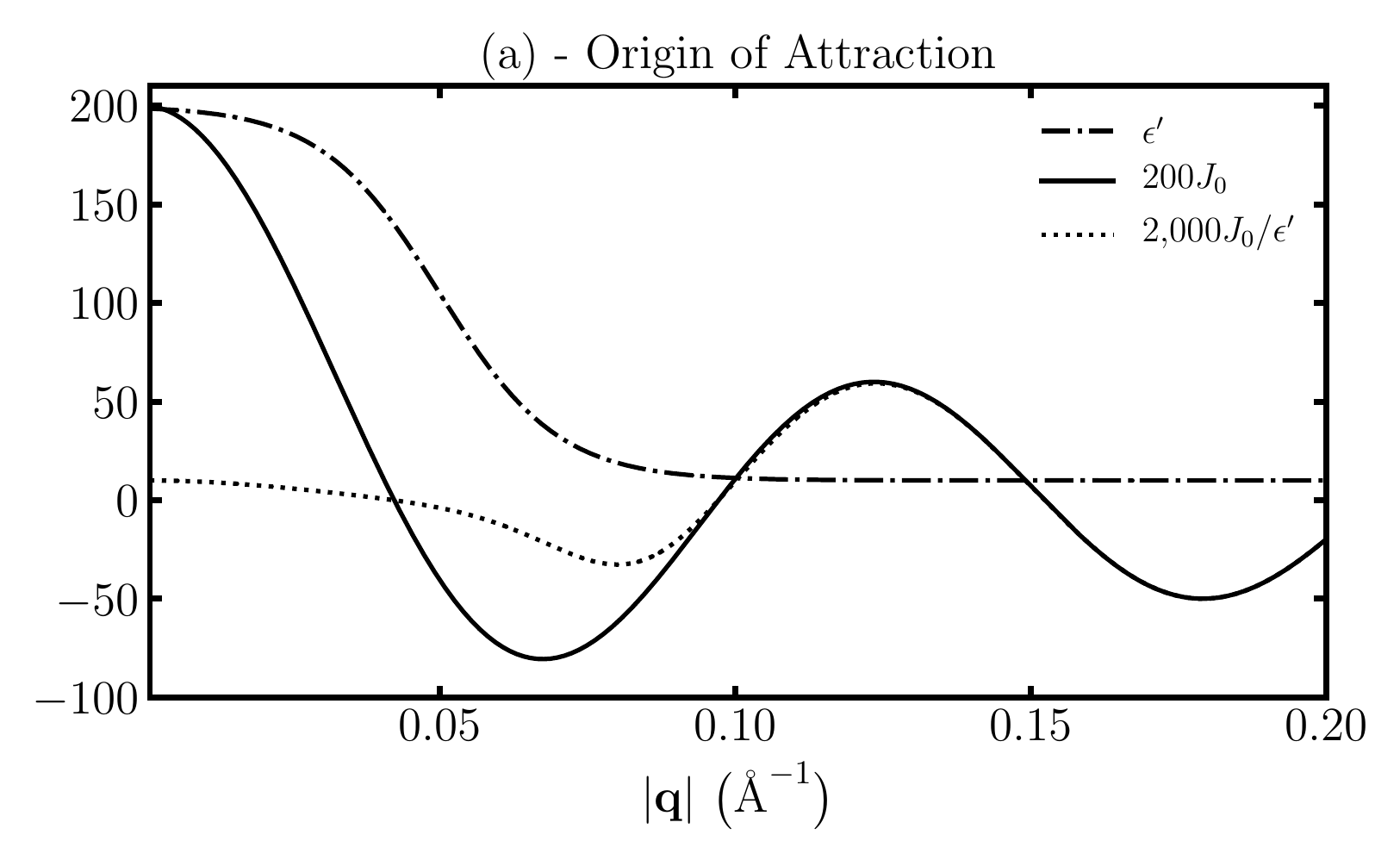}
\end{subfigure}
\begin{subfigure}{0.49\textwidth}
  \centering
  \includegraphics[width=1\linewidth]{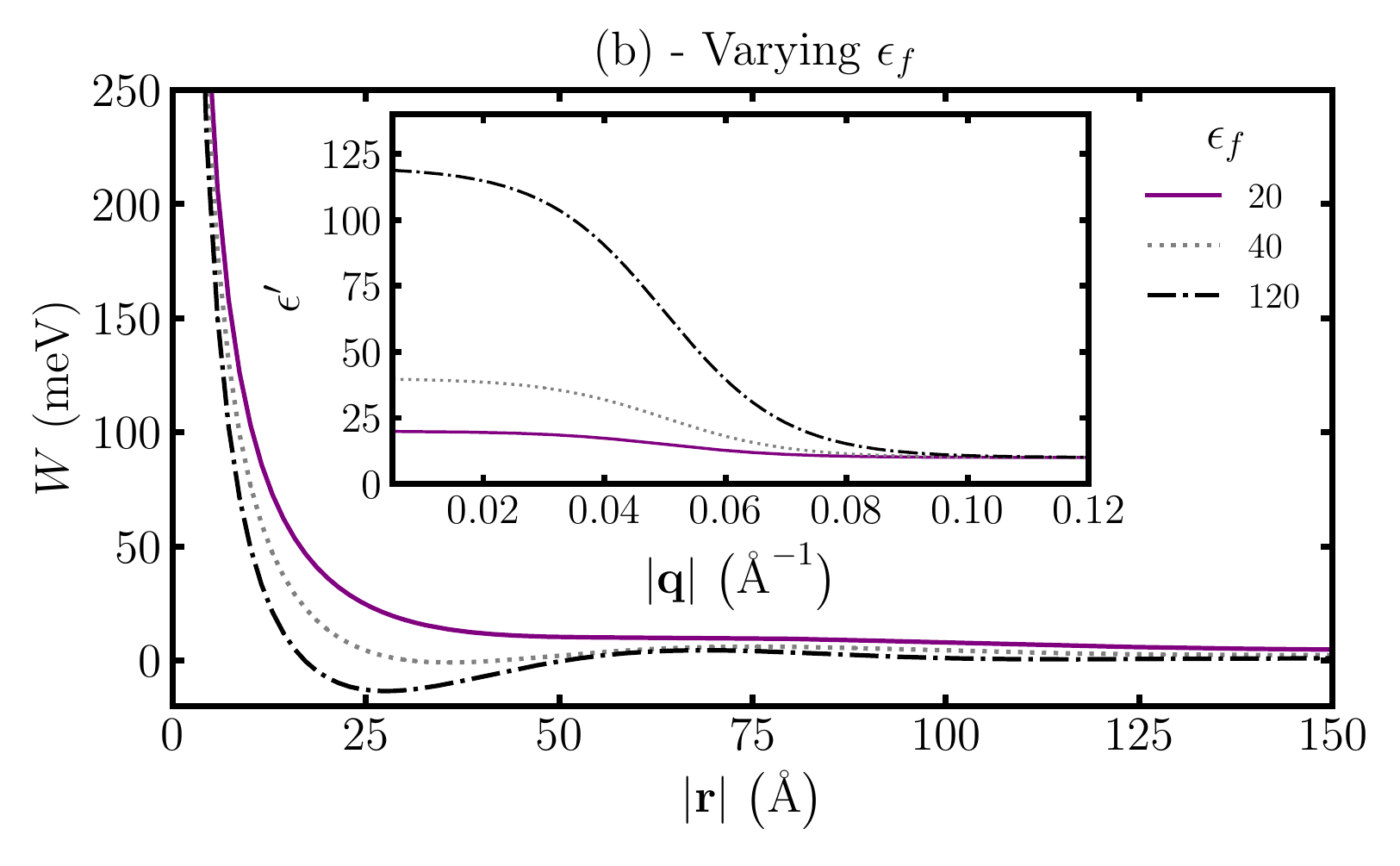}
\end{subfigure}
\\
\begin{subfigure}{0.49\textwidth}
  \centering
  \includegraphics[width=1\linewidth]{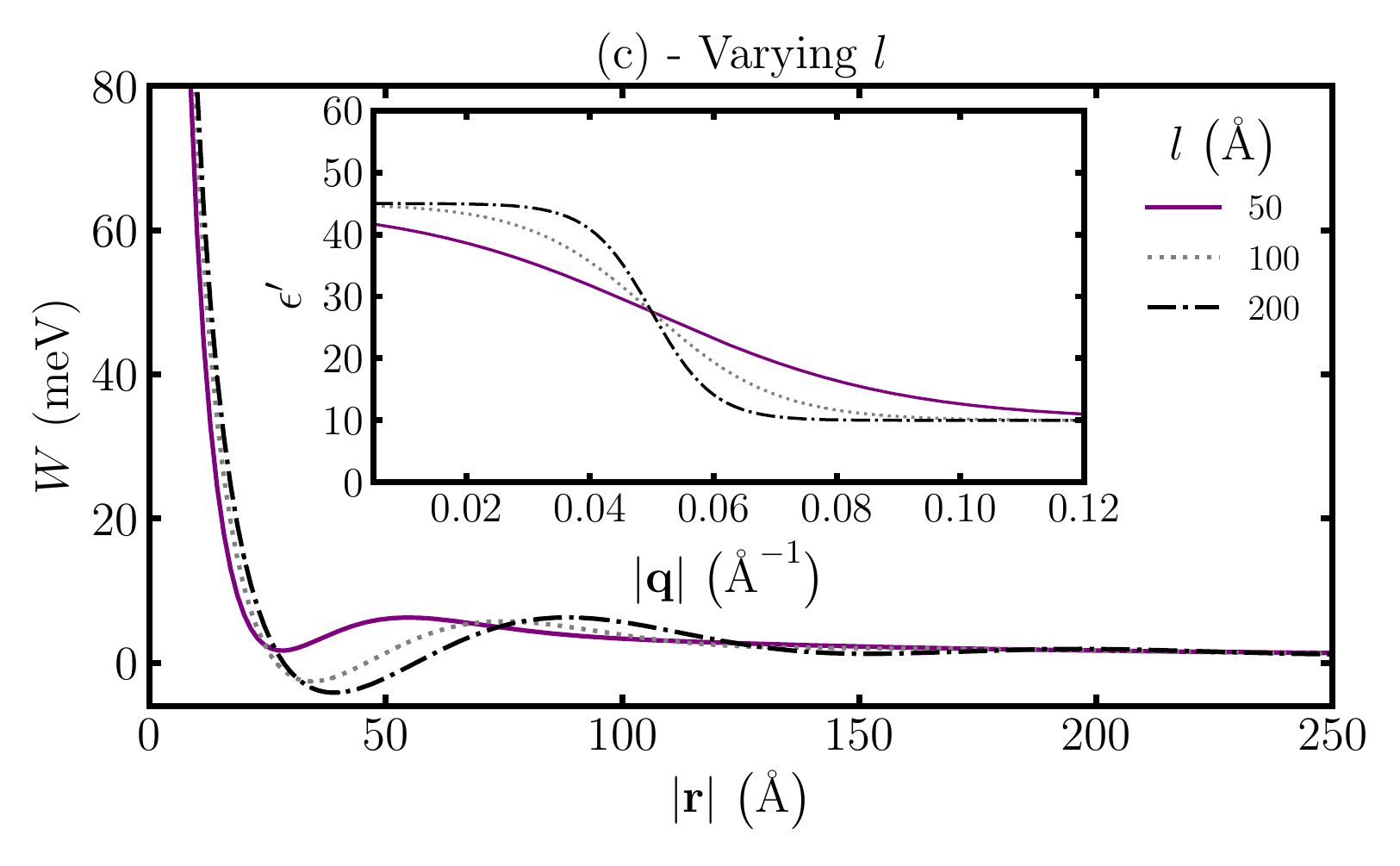}
\end{subfigure}
\begin{subfigure}{0.49\textwidth}
  \centering
  \includegraphics[width=1\linewidth]{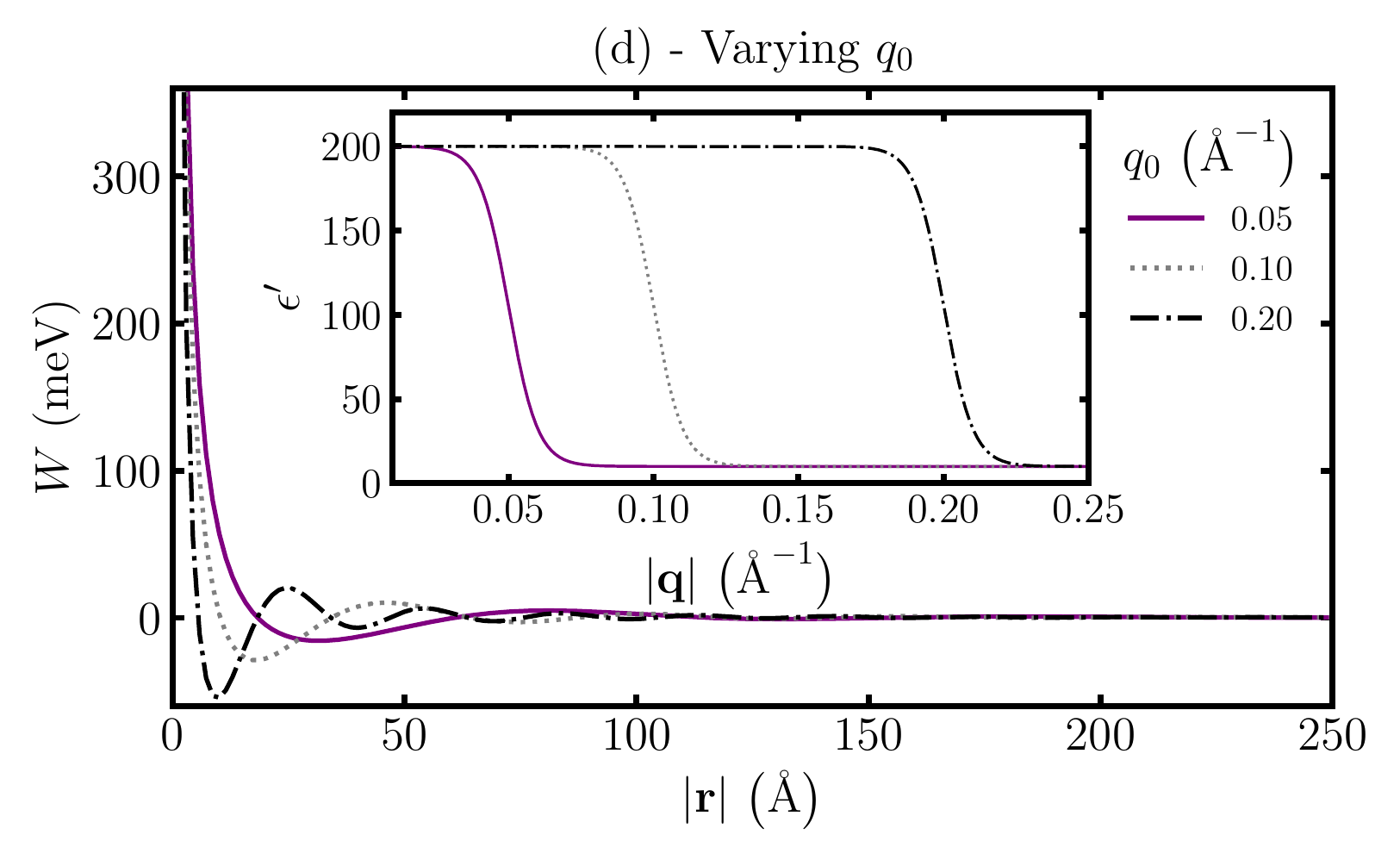}
\end{subfigure}
\caption{(a): Suppression of the small wave vector contributions of the Bessel function to the Fourier transform of the screened interaction giving rise to attractive regions in real space. (b): Screened interaction as a function of distance from the model dielectric function for several values of $\epsilon_{f}$. For the other parameters, we use $q_{0} = 0.05~\textrm{\AA}^{-1}$, $l = 100~\textrm{\AA}$ and $\epsilon_{\text{ni}} = 10$. (c): Screened interaction as a function of distance from the model dielectric function for several values of $l$. For the other parameters, we use $q_{0} = 0.05~\textrm{\AA}^{-1}$, $\epsilon_{f} = 45$, and $\epsilon_{\text{ni}} = 10$. (d): Screened interaction as a function of distance from the model dielectric function for several values of $q_{0}$. For the other parameters, we use $l = 200~\textrm{\AA}$, $\epsilon_{f} = 45$, and $\epsilon_{\text{ni}} = 10$.}
\label{TMDF}
\end{figure*}

Note that because of the highly oscillatory Bessel function, both $W_{\text{s}}$ and $W_{\text{l}}$ can potentially be negative. The RPA dielectric function, as displayed in Fig.~\ref{POL_W}(c) of the main text, goes from being a large constant to a relatively small constant within two moir\'e reciprocal lattice vectors. This dielectric function suppress the contributions of the Bessel function where it is most positive (i.e., at small values of $y$), which can give rise to negative values of the Fourier transform. This analysis is shown graphically in Fig.~\ref{TMDF}(a).

To understand the origin of the attractive interaction further, we study a model dielectric function, $\epsilon^{\prime}(q)$, with a number of free parameters. This model dielectric function must cross over from a large constant value at small wavevectors, which is determined by the Fermi velocity of the flat bands, to a smaller constant at larger momenta which is determined by the Fermi velocity of graphene. One function that meets these criteria is 
\begin{equation}
\epsilon^{\prime}(q) = \epsilon_{\text{ni}} + \dfrac{\epsilon_{f} - \epsilon_{\text{ni}}}{1 + e^{l(q - q_{0})}},
\end{equation}

\noindent where $\epsilon_{f}$, $l$ and $q_{0}$ are, respectively, the dielectric constant in the long wavelength limit, the decay length of the dielectric function and the location of the transition between these two regimes. These parameters can be varied to understand what characteristics of the dielectric function are important in giving rise to attractive regions in the effective interaction.

As can be seen in Fig.~\ref{POL_W}(c), not all twist angles exhibit attractive regions. This observation is reproduced by the model when $\epsilon_{f}$ is varied, see Fig.~\ref{TMDF}(b). For large enough values of $\epsilon_{f}$, attractive regions in the effective interaction are obtained. Another important parameter is the decay length, $l$. If the decay length is too small, the dielectric function is a slowly varying function and does not give rise to attractive regions, see Fig.~\ref{TMDF}(c). In the limit of the decay length tending towards zero, a constant dielectric factor is obtained, which gives rise to a Coulomb potential divided by a constant, i.e., there are no attractive parts of the interaction. Attractive regions are only observed for large enough values of the decay length which cause the dielectric function to vary significantly on a similar length scale of the moir\'e reciprocal lattice vector. For values of $\epsilon_{f}$ and $l$ that give rise to attractive regions in the potential, the parameter $q_{0}$ determines the period of the oscillations in the effective interaction, see Fig.~\ref{TMDF}(d).  

What happens to the attractive regions when tBLG is doped by additional electrons or holes? To model metallic systems (including tBLG at the special twist angles where the undoped system is not a semimetal, but a metal), we construct a model dielectric function that diverges at small wave vectors. This can be achieved by multiplying the second term of $\epsilon'(q)$ by $a/q$, where $a$ is some constant factor. Introducing this divergent part causes the magnitude of the oscillations to increase. Therefore, it is likely that attractive regions in the effective interaction can also be found in metallic tBLG.

\section{Wannier Functions of Flat Bands}
\label{AWF}

\renewcommand{\thefigure}{C\arabic{figure}}
\setcounter{figure}{0} 

Wannier functions for isolated band manifolds can be generated from the Bloch states via~\cite{MAVAN,MLWF}
\begin{equation}
w_{n\textbf{R}}(\textbf{r}) = \dfrac{1}{\sqrt{N}}\sum_{m\textbf{k}}e^{-i\textbf{k}\cdot\textbf{R}}U_{nm\textbf{k}}\psi_{m\textbf{k}}(\textbf{r}).
\label{eq:wannier}
\end{equation}

\noindent Here, the band index $m$ is over the flat bands only, since they are separated by energy gaps from all other bands in the whole Brillouin zone, and $N = 30 \times 30$ is the number of $k$-points utilised in the discrete Fourier transform. The unitary matrix, $U_{nm\textbf{k}}$, which mixes bands, represents the gauge freedom of the Bloch states and is determined by the Wannier90 code~\cite{W90vT} such that the resulting Wannier functions are maximally localized~\cite{MAVAN,MLWF}.

\begin{figure*}[t!]
\begin{center}
\includegraphics[width = 0.8\linewidth]{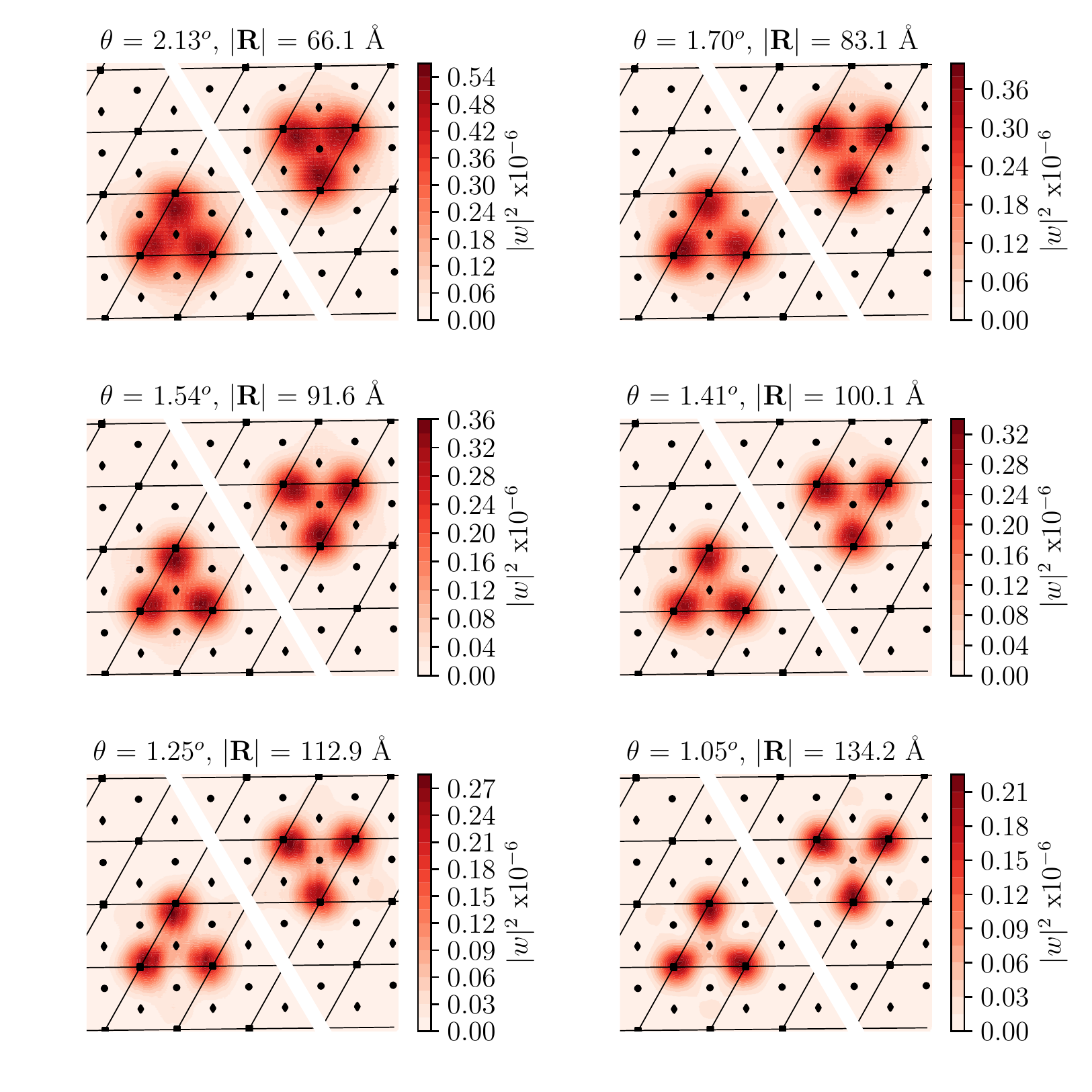}\\
\caption{Flat-band Wannier functions of tBLG for the studied twist angles. Subfigures for the twist angles of 2.13$\degree$ and 1.05$\degree$ adapted with permission from Ref.~\citenum{PHD_1}. Copyrighted by the American Physical Society.}
\label{VARF}
\end{center}
\end{figure*}

To obtain a Wannier-transformed Hamiltonian that reproduces the correct degeneracies of the band structure of tBLG, the Wannier functions must be centered either at the AB or the BA positions of the moir\'e unit cell~\cite{MLWO,SMLWF,OMIB,MMIT} (denoted by diamonds and circles in Fig.~\ref{VARF}). To achieve this, we use the approach of Ref.~\citenum{SLWF}, where a procedure to calculate a sub-set of maximally localized Wannier functions with constrained centres was outlined. This approach was chosen because it has been empirically found to produce the correct symmetries of Wannier functions, provided the center of the Wannier function is enforced at the correct position~\cite{SLWF}. Here, we constrain the centers of two Wannier functions, one on the AB and one on the BA positions, and minimize the cost function 
\begin{equation}
\Omega = \sum_{n=1}^{J^{\prime}}\Big[\braket{r^{2}}_{n} - \bar{\textbf{r}}_{n}^{2} + \lambda(\bar{\textbf{r}}_{n} - \textbf{r}_{0n})^{2}\Big].
\end{equation}

\noindent In this equation, the first two terms describe the quadratic spread of the Wannier functions, with $\braket{r^{2}}_{n} = \braket{w_{n\textbf{R}}|r^{2}|w_{n\textbf{R}}}$ and $\bar{\textbf{r}}_{n} = \braket{w_{n\textbf{R}}|\textbf{r}|w_{n\textbf{R}}}$~\cite{MAVAN,MLWF}. The third term introduces a penalty if the center of the $n$-th Wannier function deviates from $\textbf{r}_{0n}$. In our calculations, we use a value of $\lambda=200$ for the cost parameter.

This selective localization procedure is implemented in the Wannier90 code (version 3.0)~\cite{W90vT}, which requires $M_{mn}^{\textbf{k,q}} = \braket{u_{m\textbf{k}}|u_{n\textbf{k}+\textbf{q}}}$ and $A_{mn}^{\textbf{k}} = \braket{\psi_{m\textbf{k}}|g_{n}}$ to maximally localize the initial guess, $g_{n}$. Here $u_{n\textbf{k}}$ is the unit cell periodic part of the Bloch state, $\psi_{n\textbf{k}} = e^{i\textbf{k}\cdot\textbf{r}}u_{n\textbf{k}}(\textbf{r})$. Inserting the expression for the tight-binding Bloch states and retaining only contributions from pairs of $p_z$-orbitals on the same atom yields
\begin{equation}
M_{mn}^{\textbf{k,q}} = \sum_{j}c^{*}_{m\textbf{k}i}c_{n\textbf{k}+\textbf{q}i}e^{i\textbf{q}\textbf{t$_{j}$}}I(\textbf{q}).
\end{equation}

\noindent Here, we utilise the initial guess for the Wannier states from Ref.~\citenum{SMLWF}. Specifically, the initial guess is obtained by constructing a linear combination of the degenerate Bloch eigenstates at $\Gamma$ to create a new set of smooth Bloch states. These Bloch functions were then mixed such that the electron density in the vicinity of a specific AB or BA position (corresponding to the center $\textbf{r}_{0n}$) is maximized. Applying a Gaussian cut-off to this superposition, $f(\textbf{r} - \textbf{r}_{0n})$, yields a localized initial guess given by 
\begin{equation}
\ket{g_{n}} = \sum_{n^{\prime}}\psi^{v}_{n^{\prime}\Gamma}(\textbf{r})f(\textbf{r} - \textbf{r}_{0n}),
\end{equation}
\noindent where $v$ denotes the layer and sub-lattice degrees of freedom. The decay length of the Gaussian was chosen to be 0.7 times the moir\'e lattice constant.

Inserting this initial guess, we find
\begin{multline}
A_{mn}^{\textbf{k}} = \dfrac{1}{N}\sum_{n^{\prime}}\sum_{\textbf{R}\textbf{R}^{\prime}}\sum_{jv_{i}}c^{*}_{m\textbf{k}j}c_{n^{\prime}\Gamma v_{i}}e^{-i\textbf{k}\cdot\textbf{R}} \times \\ \int d\textbf{r}\phi^{*}(\textbf{r} - \textbf{t}_{j} - \textbf{R})f(\textbf{r} - \textbf{r}_{0n})\phi(\textbf{r} - \textbf{t}_{v_{i}} - \textbf{R}^{\prime}).
\end{multline}

\noindent Note that $v_{i}$ only runs over the atoms located on the layer and sub-lattice corresponding to $v$. Retaining contributions from pairs of $p_{z}$-orbitals on the same atom and using the fact that the Gaussian is a slowly varying function and be taken outside the integral, we arrive at
\begin{equation}
A_{mn}^{\textbf{k}} = \dfrac{1}{N}\sum_{n^{\prime}}\sum_{\textbf{R}}\sum_{v_{i}}c^{*}_{m\textbf{k}v_{i}}c_{n^{\prime}\Gamma v_{i}}e^{-i\textbf{k}\cdot\textbf{R}}f(\textbf{t}_{v_{i}} + \textbf{R} - \textbf{r}_{0n}).
\end{equation}

\noindent The $\textbf{R}$ summation is performed over the entire crystal. 

In agreement with the empirical observation that selectively localized Wannier functions exhibit the correct symmetry~\cite{SLWF}, we found that the Wannier functions of the flat bands exhibit of three lobes located on the AA regions~\cite{MLWO,SMLWF}. In Fig.~\ref{VARF}, the calculated Wannier orbitals for the twist angles studied in the main text are displayed.

\section{Coulomb Matrix Elements}
\label{ACM}

\renewcommand{\thetable}{D\arabic{table}}
\setcounter{table}{0} 

In a Wannier basis, the interacting contribution to the Hamiltonian is given by
\begin{equation}
\hat{\mathcal{H}}_{int} = \dfrac{1}{2}\sum_{\{n_{i}\textbf{R}_{i}\}}V_{\{n_{i}\textbf{R}_{i}\}}\hat{c}^{\dagger}_{n_{4}\textbf{R}_{4}}\hat{c}^{\dagger}_{n_{3}\textbf{R}_{3}}\hat{c}_{n_{2}\textbf{R}_{2}}\hat{c}_{n_{1}\textbf{R}_{1}},
\end{equation}

\noindent where the creation (annihilation) operator $\hat{c}^{\dagger}_{n\textbf{R}}$ ($\hat{c}_{n\textbf{R}}$) creates (destroys) an electron in the Wannier state $\ket{w_{n\textbf{R}}}$, and $V_{\{n_{i}\textbf{R}_{i}\}}$ denotes the matrix element of the screened interaction. Here, we focus on the calculations of the Hubbard parameters, i.e. the special case of $\textbf{R}_{4} = \textbf{R}_{1}$, $\textbf{R}_{3} = \textbf{R}_{2}$, $n_4=n_1$ and $n_3=n_2$. 

To evaluate Eq.~\eqref{GHI} for the screened interaction and calculated Wannier functions, the integral was re-expressed as a sum of interacting $p_{z}$-orbitals. To obtain this, the Bloch states were inserted into the Wannier functions, such that the Wannier functions are a linear combination of $p_z$-orbitals according to
\begin{equation}
w_{n\textbf{R}}(\textbf{r}) = \sum_{j\textbf{R}^{\prime}}c_{n\textbf{R}\textbf{R}^{\prime}j}\phi_{z}(\textbf{r} - \textbf{t$_{j}$} - \textbf{R}^{\prime}),
\label{WFLBS}
\end{equation}

\noindent where
\begin{equation}
c_{n\textbf{R}\textbf{R}^{\prime}j} = \dfrac{1}{N}\sum_{m\textbf{k}}U_{nm}^{(\textbf{k)}}e^{i\textbf{k}(\textbf{R}^{\prime} - \textbf{R})}c_{m\textbf{k}j}.
\end{equation}

\noindent Inserting Eq.~\eqref{WFLBS} into Eq.~\eqref{GHI} yields 
\begin{equation}
V_{n_{1}\textbf{R}_{1}n_{2}\textbf{R}_{2}} = \sum_{\textbf{R}^{\prime}\textbf{R}^{\prime\prime}}\sum_{lj} |c_{n_{1}\textbf{R}_{1}l\textbf{R}^{\prime}}|^{2}|c_{n_{2}\textbf{R}_{2}j\textbf{R}^{\prime\prime}}|^{2}v_{l\textbf{R}^{\prime}j\textbf{R}^{\prime\prime}}.
\label{FE}
\end{equation}

\noindent Here $v_{l\textbf{R}^{\prime}j\textbf{R}^{\prime\prime}}$ denotes the atomic Hubbard parameter of $p_{z}$-orbitals located on the carbon atoms with labels $l\textbf{R}^{\prime}$ and $j\textbf{R}^{\prime\prime}$ in tBLG. When the $p_{z}$-orbitals are sufficiently separated (i.e., when they are not located on the same carbon atom), a point-like interaction was assumed. When the two $p_z$-orbitals are on the same atom, we utilize an atomic on-site Hubbard parameter from DFT~\cite{SECI}. As seen in Fig.~\ref{VARF}, the Wannier orbitals are not located in a single unit cell, so it is essential that the summation is performed over a large enough supercell. We find that a $5\times5$ supercell yields converged results.

\subsection{cRPA Ohno Potential Fits}

Table~\ref{OHNOT}shows the exponents, $n$, of the generalised Ohno potential~\cite{OHNO}
\begin{equation}
V(r) = \dfrac{V_{00}}{\sqrt[n]{1 + (V_{00}/W_{\textrm{env}}(r))^{n}}},
\end{equation}

\noindent which describes the extended cRPA Hubbard parameters. As the twist angle decreases, the extended Hubbard parameters reduce to the bare Coulomb interaction between centres at larger separations. Therefore, the exponent of the generalised Ohno potential is smaller for smaller twist angles.

\begin{table}[htp]
\begin{center}
\begin{tabular}{lcr}
\hline\hline
$\theta$ / degree & $\quad\quad$ & $n$ \\
\hline 
2.13 & & 2.5 \\
1.70 & & 2.4 \\
1.54 & & 2.2 \\
1.41 & & 2.0 \\
1.25 & & 1.3 \\
1.05 & & 1.1 \\
\hline\hline
\end{tabular}
\end{center}
\caption{Generalised Ohno potential exponents for extended cRPA Hubbard parameters.}
\label{OHNOT}
\end{table}

\bibliographystyle{apsrev4-1}
\bibliography{ADI}

\end{document}